# Engaging young people for a more inclusive national energy transition: A participatory modelling framework

Muhammad Shahzad Javed[1,*], Karin Fossheim[2], Maximilian Roithner[1], Matylda N. Guzik[1], James Price[3], Beate Seibt[2], Marianne Zeyringer[1]

[1]Department of Technology Systems, University of Oslo, Norway.

[2]Department of Psychology, University of Oslo, Norway.

[3]UCL Energy Institute, University College London, United Kingdom.

[*]Corresponding author(s). E-mail(s): m.s.javed@its.uio.no

## Abstract

Participatory research in energy system modelling can generate bottom-up knowledge to explore co-designed future net-zero energy system scenarios. However, it often fails to facilitate collective learning, explore explicitly informed perspectives, and frequently ignores underrepresented groups like youth, among whom distrust about the energy transformation process is growing. By modifying a national electricity system model to reflect young people's socio-techno-environmental insights gathered through school workshops, this study presents a framework for envisioning future net-zero power systems in Norway. Given pupil priorities regarding certain power system aspects and their cumulative impact, substantial shifts occur in national renewable capacity potentials (approximately ±50%), system costs (-7% to +25%), technology mixes (notably onshore wind from 40% to 0%), transmission capacities (near doubling), and regional equity assessments. We find that costly youth-driven system designs do not necessarily guarantee equitable systems. Although applied to young people in Norway, the proposed workshop-informed modelling framework serves as a tool to meaningfully engage diverse groups and capture their perspectives, thereby further democratising energy system planning. The approach is expected to help address social acceptance challenges through enhanced understanding of trade-offs in the energy transformation process.

## 1. Introduction

There is a growing consensus on the necessity of interdisciplinary approaches to analyze and design future energy systems. One area where this requirement manifests is in siting decisions for new renewable energy technologies (RET), often perceived as posing social and environmental risks and injustices [1]. In particular, the visual impact, changes in landscape aesthetics and environment, and unequal distribution of costs and benefits pose considerable barriers to accepting new RET deployments [2,3]. These concerns can erode trust in public institutions and democratic processes, prompting governments to retract or reconsider proposed RET developments [4–6]. One example is the abandonment of the Norwegian Water Resources and Energy Directorate (NVE) national framework proposal, which intended to pinpoint viable areas for onshore wind development spanning 29,000 $km^2$ (9% of total land area), with an estimated capacity of ~290 GW, due to significant public criticism [7]. In response, governments increasingly decentralize the RET planning process to ensure projects offer socio-economic benefits to all community groups [8,9].

Since 2019, underrepresented groups, particularly young people, have increasingly voiced concerns on energy and climate issues through climate strikes, protests, and legal actions [10], advocating for meaningful measures through filing climate justice cases and a more inclusive energy transition process [11]. For example, Norway-based youth-led groups like "Natur og ungdom" ("Nature and Youth") and "Fremtiden i våre hender" ("The Future in Our Hands") promote nature preservation and have been opposing wind power installations for this reason. Monetary considerations cannot fully address the concerns and values young people hold regarding the energy transition [12]. Studies indicate that young individuals promote environmental literacy and are eager to contribute to the energy transformation [13]. Despite growing evidence of climate-related psychological distress among youth worldwide [14], there remains a lack of large-scale, meaningful engagement with them. Thus, empowering youth can help address the overlooked concerns of marginalized groups in the energy transition, thereby enhancing social acceptance through their forward-looking stance, willingness to engage with new information [15], and involvement in internet-based social movements [16].

Energy system modelling (ESM) is an essential tool for studying the design of future energy systems that meet commitments under the Paris Agreement. Their techno-economic nature provides insights into challenges, transition pathways, and sensitivity to technological and policy developments. Despite their role in supporting decision-makers, quantitative ESMs often fail to sufficiently integrate stakeholder perspectives necessary to adequately address the energy transformation's social dimension alongside techno-economic details [17]. Participatory approaches in ESM enable the design of scenarios by generating bottom-up knowledge of stakeholder perspectives regarding energy transition [18,19]. Participatory ESM have used approaches like online surveys [20], factsheets [21], group discussions and interviews [22], multi-criteria decision making [23], and interactive tools [24–26] (**Supplementary Table 1)**. Although approaches like interactive tools facilitate learning about energy system planning by exploring various combinations of system components, they often fail to support collective learning. This limits their ability to address knowledge gaps about renewable electricity technologies and provide context for informed energy choices, potentially leading to misconceptions about impacts [3,18]. Moreover, explicit stakeholders' perspectives on power system components, particularly regarding acceptance and equity, remain relatively unexplored [27]. For example, participants might choose onshore wind under conditions of excluding certain landscapes and ensuring regional equity. Furthermore, most participants in previous studies consist of various experts, with the general public comprising fewer than 20% of available studies, indicating certain groups, especially young people—except for study [24]—are overlooked [18,27–29]. Consequently, a framework meaningfully incorporating particular stakeholder groups across the ESM process—including scenario design, optimization, results discussions, and feedback— remains lacking [28–31].

One way to address this is through educational workshops during school hours, which may act as a building block for fostering mutual understanding between young people, local stakeholders, and national planners. Such initiatives, developed as intervention tools, can enable young people to form and express opinions about the energy transformation process. We conduct full-day compulsory workshops (three sessions of 90 minutes each) in five schools, involving 286 students aged 15-16 from diverse backgrounds. Through interactive activities and games during these sessions, we discuss the basics of the green energy transformation, potential conflicts related to RET installations, and climate justice. These Workshops facilitate collective learning and provide context for informed energy choices, aligning with our aim of empowering and meaningfully engaging young people in the energy system planning. We design questionnaires to collect student perspectives on a number of key power system aspects. We modify a spatiotemporally detailed electricity system planning model [32], incorporating various settings based on students' preferences. These include technology choices, regional preferences on the RETs siting, self-sufficiency, choices for new power lines build out, and the exclusion of certain landscapes, while ensuring an adequate net-zero power system. (**see Methods**, **Supplementary Note 2, Supplementary Table 6**). Although systematically integrating stakeholder perspectives may result in higher energy system costs, modelled scenarios may perform better on non-modelled objectives such as political feasibility, social acceptance, and energy justice [33,34]. This aligns the ESM more closely with real-world considerations and political discourse.

## 2. School workshop-informed modelling framework

The scenarios based on students' input capture power system aspects that constitute most of the system cost and encompass land use conflicts and visual impacts; factors that challenge social acceptance [27]. The base model includes only core technical, environmental, and geographical restrictions from Ref. [35] without incorporating students' preferences. The transmission (trans) scenario reflects students' choices regarding maintaining or expanding current power lines with visible overhead or expensive underground power lines. The trade scenario represents inclinations regarding electricity trade levels with neighbouring countries, with options to decrease, increase, or maintain current import levels. The technology (tech) scenario reflects students' selection of RETs (wind and/or solar) through discrete choice experiment questions. Based on national map data, the landscape (land) scenario integrates preferences for nine common Norwegian landscape types for onshore wind development [36]. The region scenario indicates students' spatial choices for siting new RETs across Norwegian counties. We derived preference coefficients from a detailed assessment of workshop responses and integrated them into the modified energy system model.

We assess the impact of individual, cumulative, and different prioritisation sequences of student choices on future net-zero electricity system design. **Fig. 1** illustrates the framework workflow. A follow-up feedback session encouraged students to reflect on the outcomes of their energy choices by discussing how their preferences influence the design of a future net-zero electricity system. The modelling output thus enabled a meaningful and fact-based dialogue between students on possible future electricity system designs. Complete methodological details of workshops, scenarios, and energy system model modifications are described in the Methods section.

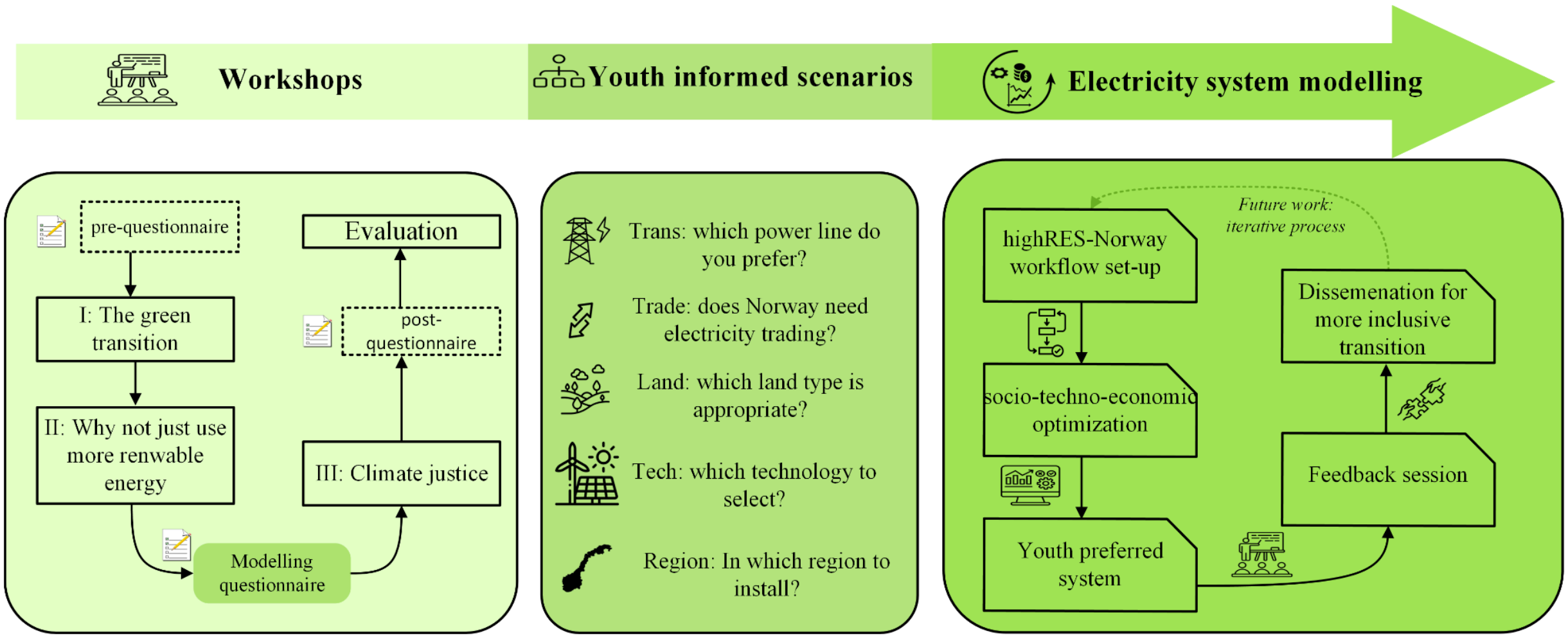

**Fig. 1** School workshop-informed multidisciplinary workflow for inclusive power system modelling to develop more socially acceptable energy transition pathways. Workshops consisted of three sessions discussing the basics of green transition, potential conflicts of only building more RET, and climate justice. The depicted questions are rephrased from the modelling questionnaire used to capture students' perspectives on various aspects of the power system. Exact wording can be found in the workshop materials.

## 3. Impact of "*excluded landscapes*" on national capacity potential

Given that debates over RET deployment often center on environmental and landscape impacts, we find that students' preferences as to which landscapes they would include or exclude from RET deployment affect Norway's renewable potential in different ways. **Fig. 2a** represents the proportion of land available for RETs relative to each county's total area after excluding the landscape types students opposed—primarily agricultural, residential, and forested areas. These landscapes are predominantly located in Southern and Eastern Norway, where 57% of the population resides. Although accounting for 42% of electricity demand, landscape preferences there lead to a 72% reduction in generation capacity potential, decreasing the country's onshore wind capacity potential from 371 to 180 GW (**Fig. 2b**). While selected wind sites align with students' societal and environmental priorities, their distance from demand centers and infrastructure **(Supplementary Fig. 15)** may necessitate expanding energy storage or transmission capacities. This introduces logistical (i.e., constructing and maintaining the grid in rugged and remote terrains) and possible acceptance challenges associated with transmission lines and incurs extra costs related to storage.

The more than twofold reduction in Norway's land-based wind capacity potential becomes concerning, given the projected rise in electricity demand by 2050 [37]—highlighted by the green areas in **Fig. 2d**—combined with an estimated 10% increase in the levelized cost of electricity (LCOE), which underscores the economic impact of the landscape preferences. Summarizing, the decreased available area for RETs leads to higher system costs due to: i) deploying capacity in suboptimal locations that are either less windy or distant from demand centers, requiring transmission expansion; ii) potential total capacity limits on onshore wind due to socio-economic restrictions limiting its national-scale contribution to decarbonization; and iii) increased reliance on more expensive alternatives such as offshore and floating offshore wind (with their own future transmission implications) alongside greater storage and flexibility requirements. While excluding unfavoured landscapes marginally reduces the country's mean onshore wind capacity factor by 3.1%, the utilization ratio (actual generation divided by theoretical generation) decreases on average by 13.78%, reflecting system-level bottlenecks of integrating landscape preferences **(Fig. 2c)**. In contrast, the solar utilization ratio decreases minimally by 0.84%, indicating its adaptability from both technical and social perspectives (**Fig. 2e**).

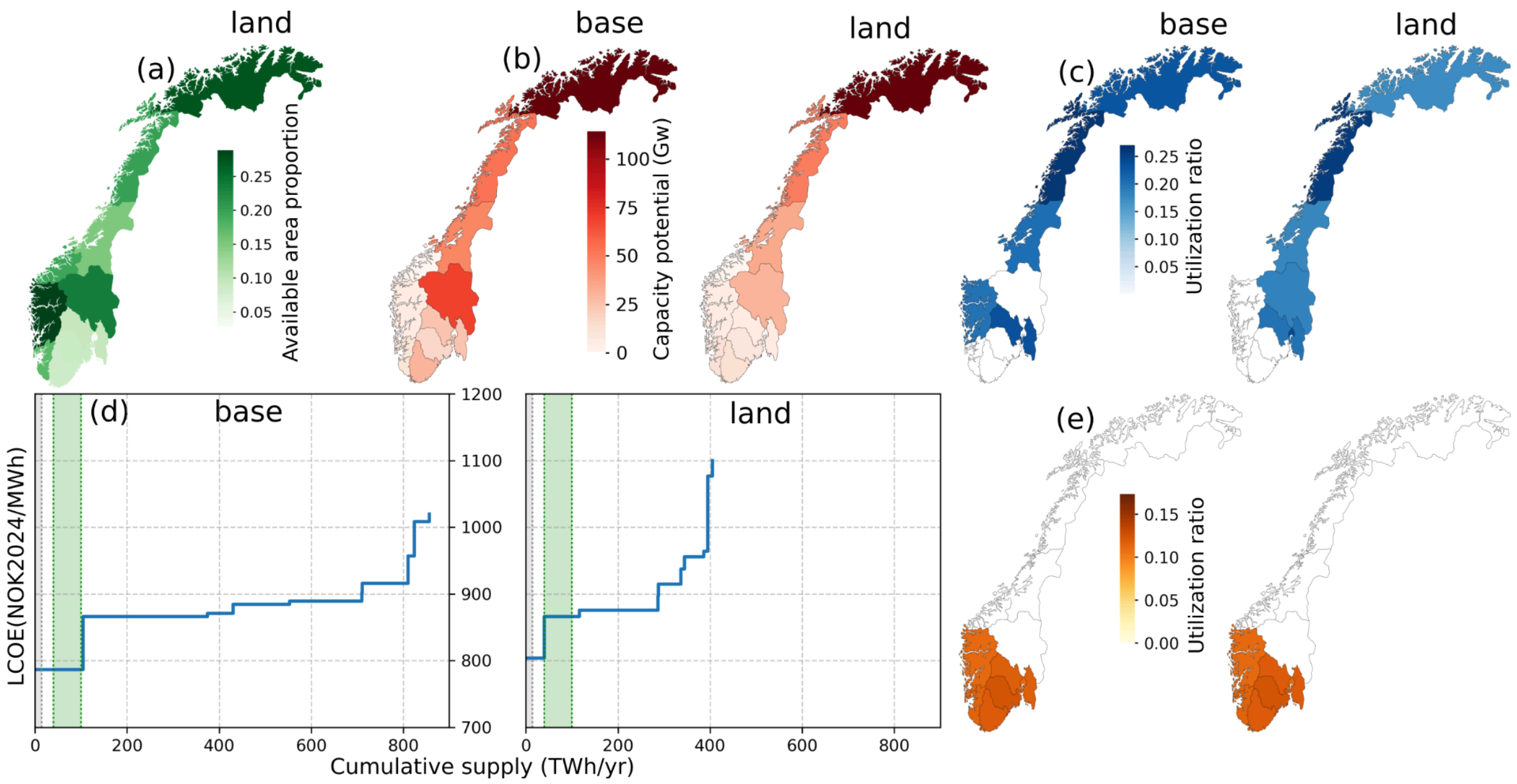

**Fig. 2** Impact of landscape choices. (a) Proportion of zonal area available after excluding opposed landscapes; (b) Comparison of the countrys' onshore wind capacity potential before (base) and after (land) excluding opposed landscapes. Available areas are aggregated into zones (Norways' administrative regions) and subsequently converted to capacity potential; (c) Utilization ratio of onshore wind (actual generation divided by theoretical generation) before (base) and after (land) the landscape exclusions; (d) Cost-potential and supply curve for land-based wind installations, with each step corresponding to a zone. The grey line represents the wind power generation in Norway in 2024, while the light green area represents the projected additional electricity demand by 2050; (e) Utilization ratio of solar before (base) and after (land) landscape exclusions, illustrating minimal change. See **Supplementary Fig. 1** for solar capacity potentials and **Supplementary Fig. 15** for zone descriptions.

## 4. Students' expectations change technology compositions and system costs

**Fig. 3** shows the national installed RET capacities for net-zero 2050 electricity systems, excluding hydropower. Deviations from the baseline scenario decrease total new RET capacities, except when landscape preferences are incorporated, which increases total capacities. This is because the model invests more in offshore wind, which has higher capacity factors and thus requires less capacity. Young people's choices substantially reduce the share of onshore wind capacity, from ~40% (base) to 0% under stringent regional considerations. This is broadly substituted by offshore wind, increasing its proportion from 9% to as much as 82% for the net-zero 2050 electricity system. Regional choices are most restrictive, reflecting students' decisions on what, where, and how much RET to install at the zonal level **(see Methods)**. Although converging to similar total capacities, fig. **3b** and **3c** illustrate the most contrasting outcomes across various prioritization sequences of the student’s choices. They showcase a quantification of the incremental implications of prioritizing certain aspects of power system for a more inclusive energy transition. Although hydropower dominates Norway’s electricity system (>85%), results indicate that additional energy storage is needed by 2050, fluctuating between 70 and 125 GWh depending on the choices made **(see Supplementary Figs. 4–6 and related discussion)**.

Several key insights emerge: Firstly, irrespective of socioeconomic preferences, the order, or the higher costs associated with offshore floating wind, it is anticipated to contribute to the future electricity system, ranging from 5% to 70%, owing to favorable wind conditions along the Norwegian coast and deep water. Secondly, the deployment of onshore wind is sensitive to local acceptance and regional perspectives. Lastly, a consistent solar deployment (20%-50%) is evident despite Norway’s limited annual sunlight, indicating the utilization of higher-capacity factor regions near demand centers, such as the South and East of Norway **(Fig. 5)**. Similar trends are observed across 2030 and 2040 demand years **(Supplementary Figs. 2-3)**.

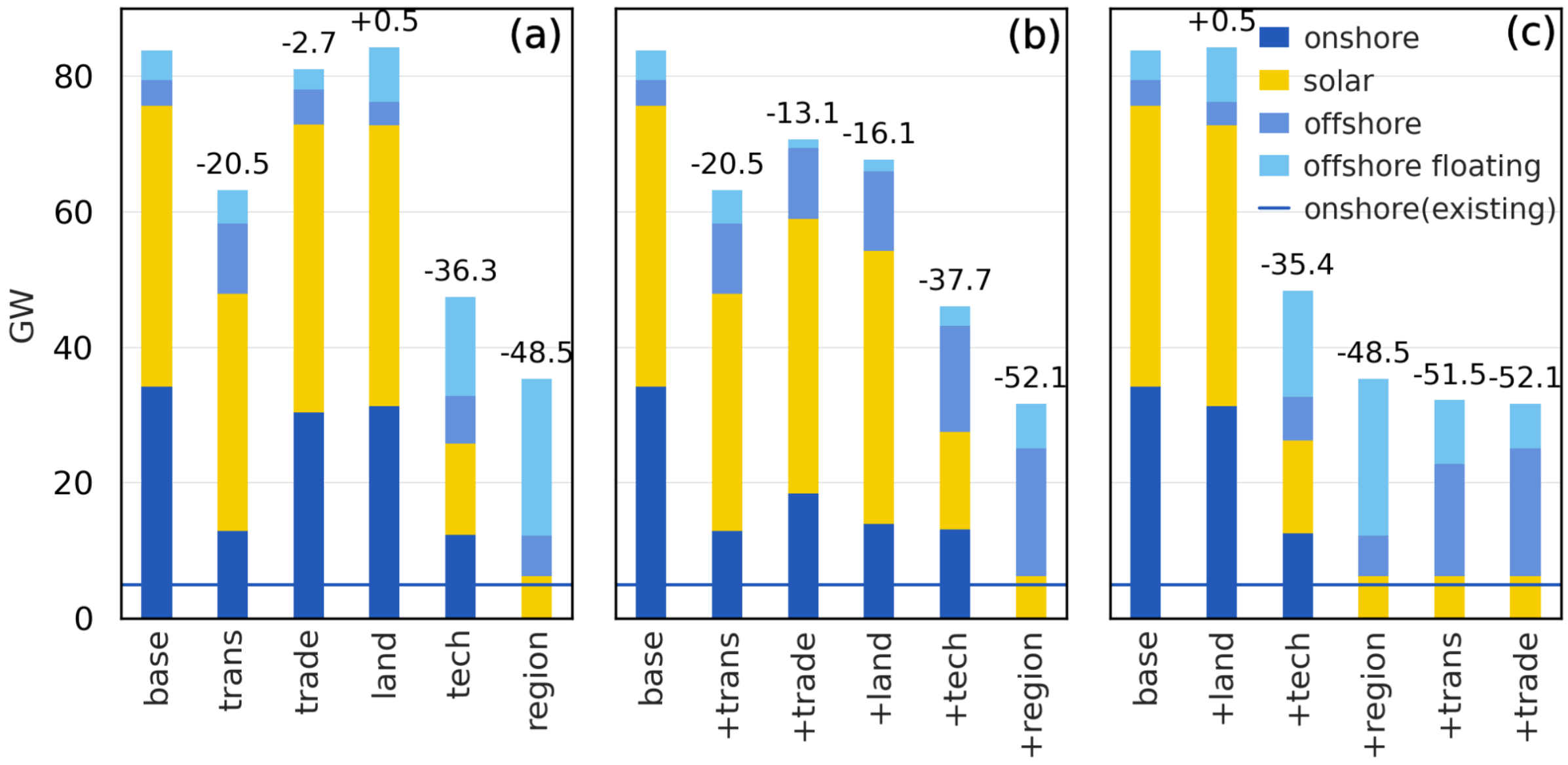

**Fig. 3** Modelled national installed capacities for a 2050 net-zero emission electricity system. Figure a demonstrates the impact of each scenario individually, while Figures b and c depict the cumulative capacities as student preferences are sequentially and incrementally added in a specified order. For context, the line indicates the current installed capacity of onshore wind in Norway. Reductions in onshore wind capacity are largely substituted by offshore wind.

Transmission and trading preferences reduce the system cost by 6-7% (blue line in **Fig. 4a**), attributed to substantial transmission infrastructure investments, which permit the deployment of renewables in more cost-effective locations for the system. Although annual electricity import and export levels are matched to reflect students' desire for self-sufficiency, increased dispatch flexibility contributes to lowering the cost. Total network capacity increases from 55 to 96 GW, facilitating connections for new offshore installations from the north and west to eastern and southern located demand centres **(Fig. 5, Supplementary Fig. 13.)**. Conversely, prioritizing landscape, technology, and regional choices increase system costs by up to ~25%; however, these costs can be offset through transmission and trade choices even under the most stringent regional priorities (**Fig. 4a, brown bars**). **Fig. 4b** indicates that, apart from the most constrained region scenario, the median cost for the 2050 net-zero system is generally lower than the baseline, regardless of the prioritization or strictness levels indicated by students' choices. This outcome results from the flexibility enabled by transmission expansion and increased cross-border electricity trade. The transmission and trade sensitivity analysis further confirms these median system cost reductions **(Supplementary Figs. 8–10)**.

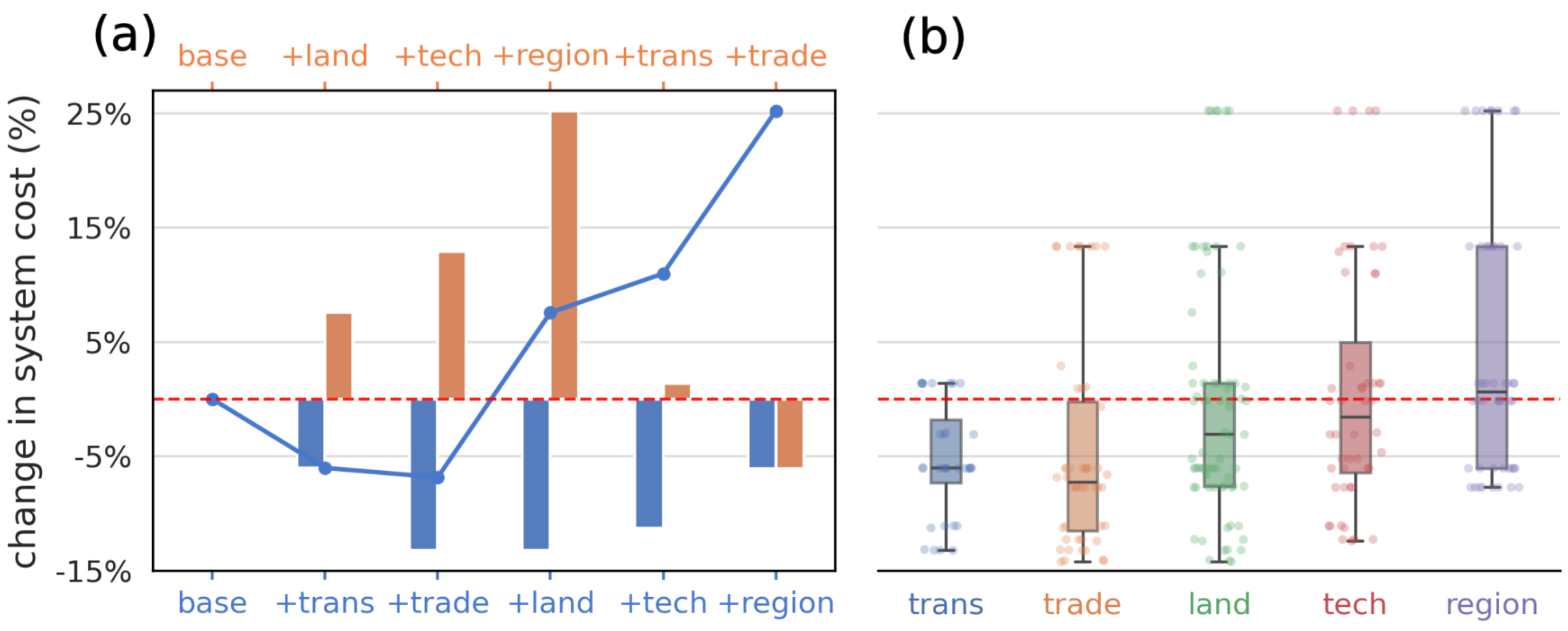

**Fig. 4** System costs for a 2050 net-zero electricity system. (a) Change in the system costs from the base scenario. The blue line indicates the impact of the individual scenarios, while the bars depict the cumulative effects of scenarios applied in the specified contrasting order. The blue color refers to the bottom x-axis, whereas the brown color corresponds to the upper x-axis. See Supplementary Fig. 7 for system costs change concerning the 2030 and 2040 demand years. (b) Sensitivity of system cost when the highest priority is assigned to a specified scenario, with other choices incrementally added following all possible sequencing orders (Supplementary Table 5). The dotted red line denotes the base scenario. System cost compositions are in **Supplementary Figs. 11–12**

## 5. Regional equity and priorities

While **Fig. 3** depicts the influence of young people's choices on the national RET capacity mix, the RETs' spatial allocation offers further insights into the impact of socioeconomic preferences on the system equity. Spatial patterns of transmission and trade preferences exhibit a more uniform distribution of solar capacities in eastern and southern Norway than baseline (**Fig. 5**). Interestingly, certain western (Møre og Romsdal, Vestland, and Rogaland) and eastern (Innlandet) zones exhibit only negligible new capacities, despite having substantial available land for RET installations and high electricity demand (**Fig. 2a, Supplementary Fig. 15**). The Norwegian government's objective to develop the majority of offshore wind farms along the western coast is primarily realized under regional priorities and technological choices. These new offshore wind installations leverage existing hydropower capacities in the west and interconnections with Germany, the Netherlands, and the UK. Although onshore wind is entirely phased out under regional preferences, unlike in other scenarios, solar installations are modestly distributed across eastern and southern zones, reflecting the exhaustion of rooftop potential adhering to preferences.

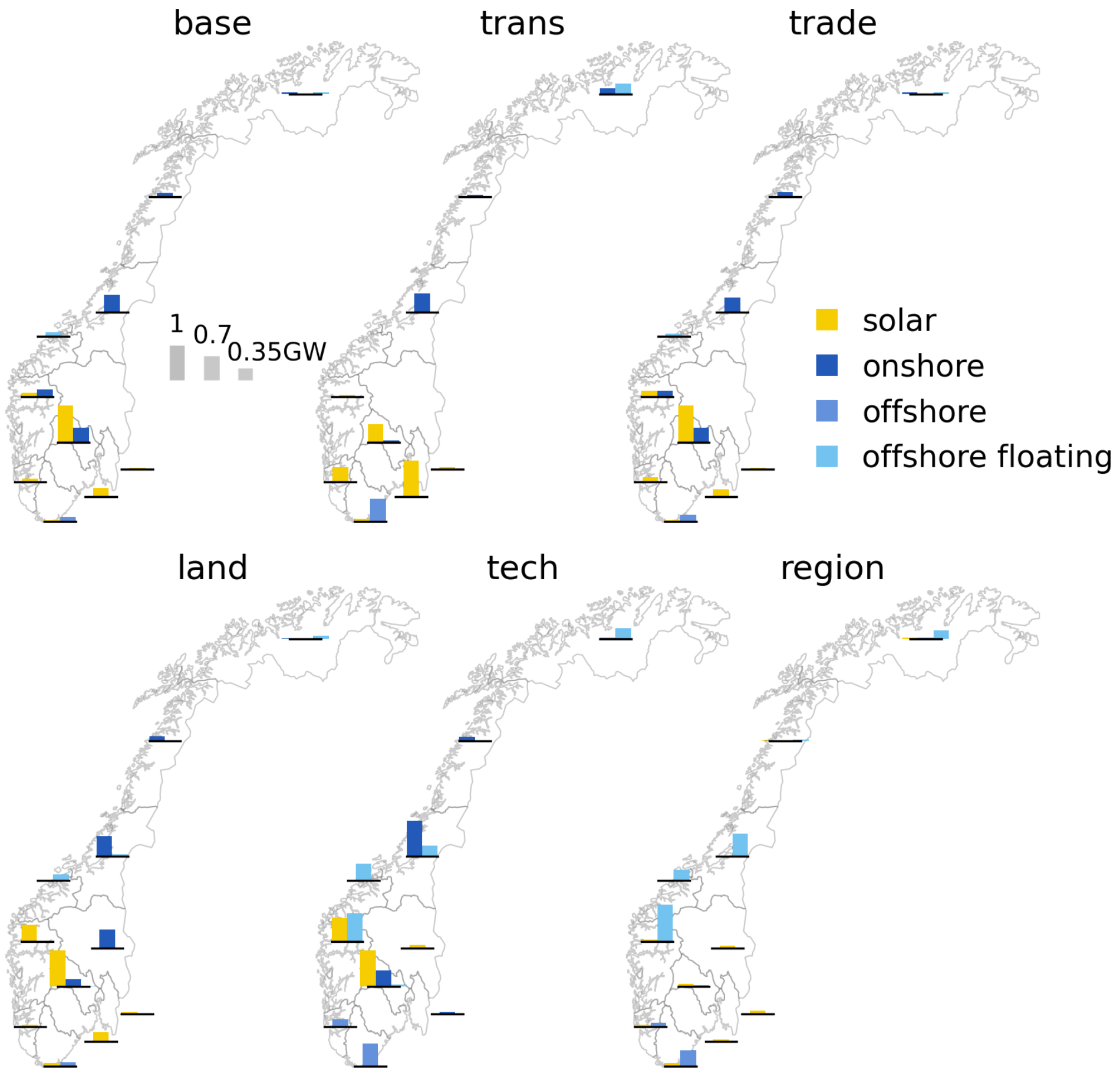

**Fig. 5** Spatial allocation of new capacity installations across zones. Capacities are normalized to the maximum capacity across technologies within the given scenario, facilitating a comparative assessment of capacities across zones. **Supplementary Fig. 13** illustrates the current spatial allocation of transmission capacities and new additions respecting student choices**.**

The obtained student-based system configurations allow us to evaluate distributional equity variations across zones, using the widely adopted statistical indicator, the Gini coefficient. Since distributional justice can be defined in multiple ways, we examine the distribution of electricity infrastructure relative to electricity demand, total land area, and population per zone [38]**(see methods)**. Across the considered distributional justice framings, the trans scenario generally emerges as the most inequitable one, while the land scenario is the most equitable, followed by the tech scenario (**Fig. 6**). The total system cost variation between the least and most inequitable scenarios is ~15%, suggesting that more equitable solutions come with moderate additional costs that must be weighed against the benefits of a just and more inclusive energy transition. Inequality arises from concentrated investments in resource-rich areas and transmitting power to distant demand centers, resulting in disproportionate infrastructure burdens across zones. Unlike the most inequitable scenarios, landscape exclusion results in a more distributed RET capacity, rather than concentration in specific zones (**Fig. 5)**.

Our analysis indicates that student-influenced net-zero power system configurations do not necessarily increase equity with increased costs. For instance, a 25% increase in system cost worsens equity by 24% and 57% under self-sufficiency and equality principles (**Fig. 4a, Fig. 6**). Conversely, a cost decrease of up to 7% under transmission and trade choices also deteriorates equity by up to 50%. **Fig. 6** reveals that different scenarios result in varying equity levels within a given justice definition and do not align across different framings of distributional justice. The heatmap (**Supplementary Fig. 14**) illustrates how the equity score shifts with changes in justice definitions. For instance, Oslo (NO03) performed best under the land-burden principle due to its small area, whereas all other zones outperformed Oslo under self-sufficiency and equality.

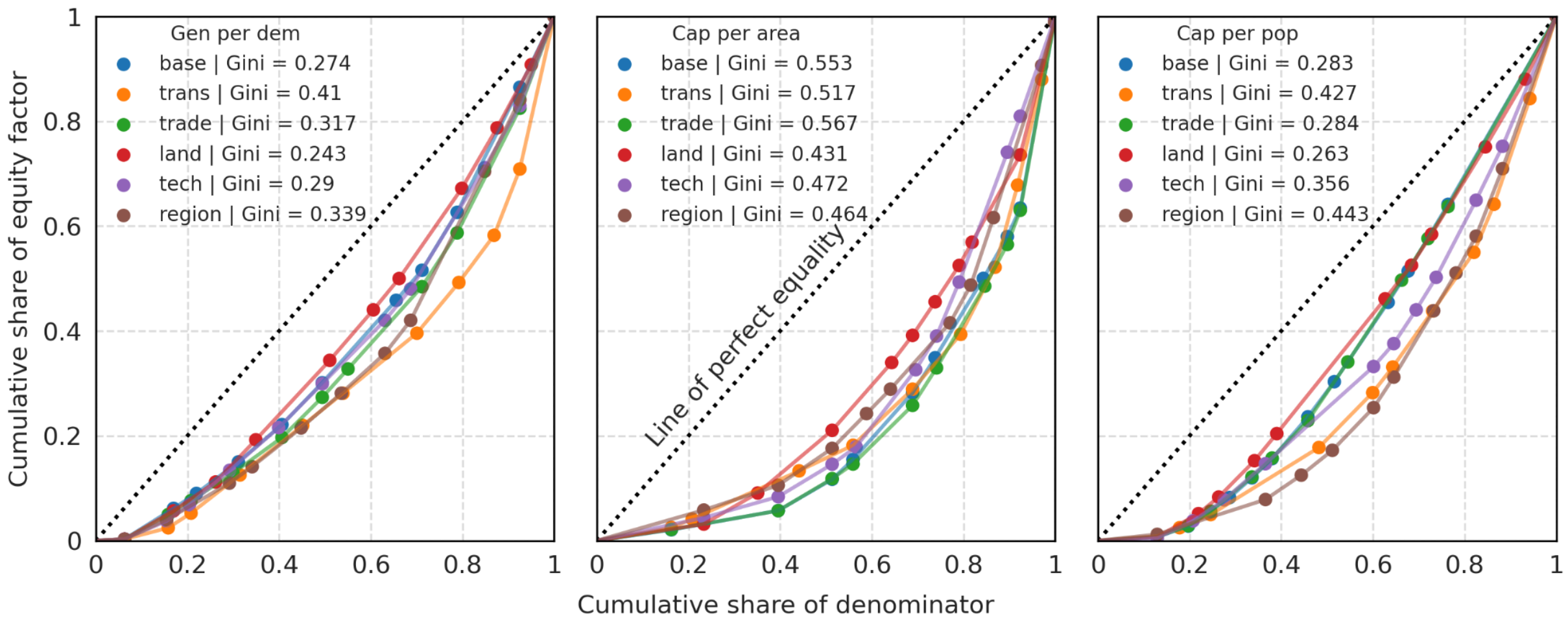

**Fig. 6** Equity assessment of student-driven scenarios for different distributional justice definitions, including self-sufficiency (generation per demand), land-burden (capacity per area), and equality (capacity per population).(**See Methods**). A Gini value close to 0 indicates high equity, whereas a value near 1 denotes high inequity. The cumulative share of the denominator represents the contribution of the relative equity factor of each zone. The figure shows the results for the 2050 electricity system design. The map plots of denominators used for Gini coefficient calculation are shown in **Supplementary Fig. 15.**

## 6. Qualitative discussion and feedback on the net-zero systems designed by students' choices

To gather feedback on the modelling outcomes reflecting participants' choices, we held a session at one of the participating schools and presented the modelling results. Students largely reaffirmed their support for offshore wind and solar power, while deepening their understanding of the trade-offs and the necessity of compromises. Students viewed a high reliance on offshore wind as a pathway to sustainability, environmental preservation, and the avoidance of localised socio-political tensions. As one participant noted, "Offshore and solar make sense to me." However, marine stakeholders' perspectives on offshore wind development are essential for building just offshore energy systems. Students were surprised by the prevalence of onshore wind in certain scenarios, with one stating, "I am surprised there is so much onshore." The reasons for their surprise were diverse, including considerations of costs, valuing nature, and social factors. While students generally adhered to the country's self-sufficiency, some desired to help other nations address climate change by increasing electricity exports.

Students foresee limited new job opportunities due to potential strains on local economies stemming from decreased reliance on oil and gas exports in the net-zero electricity systems. The high system costs associated with regional preferences for RET distribution appeared to influence students to reconsider these aspects, reflected in notes like, "...Though we have aesthetic/environmental demands, the cost is too high to consider all of these". The feedback session revealed several benefits: providing students with opportunities to reassess their choices, deepen their understanding of renewable electricity systems and associated trade-offs, and address misconceptions commonly spread through social media—for instance, that the cost of excluding wind energy development in the energy transition is not substantial, electricity trading among countries is not indispensable to achieving net-zero, and wind turbines consume more electricity than they produce.

## 7. Discussion

The global drive to reduce GHG emissions has led, and will continue to lead, to increased deployment of solar and wind installations. These often provoke economic conflicts and acceptance challenges due to visual, auditory, environmental impacts, and land use considerations. The proposed workshop-informed modelling framework provides a template to expand educational outreach and engage diverse community groups meaningfully. Our results supplement the insights from academic literature and policy documents, addressing themes such as self-sufficiency (trade scenario), nature protection (land scenario), willingness to accept/pay (technology scenario), and empowerment of underrepresented groups [18,25,27,39].

We show that excluding landscapes opposed by students could increase the LCOE by roughly 10% to meet the 2050 demand forecast for Norway due to onshore wind being replaced with solar PV and offshore wind (**Fig. 2**). Compared to a Great Britain-based study, where public sensitivity to visual impacts increased total system costs by up to 14.2%, students' preferences in Norway have a smaller effect [40]. Moreover, the results closely align with current policy discussions and the Norwegian government's plan to implement offshore wind capacity of 30GW, including 1.5GW from floating turbines by 2040. This target would offer a counterbalance to the twofold reduction in the country's onshore wind capacity potential based on our results [41].

The students' increased support for rooftop solar can be applied to areas with high solar capacity potential, such as southern and eastern Norway, to partly fill the gap of reduced onshore wind. Although solar PV upfront costs are relatively low compared to other RETs, adoption issues in Norway may stem from insufficient policy support addressing the price-production mismatch, where electricity prices are very low during the summer when solar PV output is higher and high in the winter when PV output is lower, thereby lowering incentives for rooftop solar investment.

Our insights indicate that the net-zero energy system envisioned by young people does not necessarily deviate significantly from policymakers' perspectives or techno-economic ESM scenarios [35]. This suggests youth empowerment as a critical strategy for fostering broader community support and mitigating the growing negative sentiment towards RETs. While alignment between policymakers and youth in advocating for extensive development of offshore wind installations may generally strengthen public trust and support, it also risks generating localized socio-political tensions due to regional inequities and visual and environmental impacts on coastal communities. These installations would be concentrated in western and southern Norway, where hydropower resources already exist and export to demand centres **(Fig. 5 & Fig. 6)**.

The open question of how equitable the resulting unified net-zero system will be is largely influenced by the definitions of distributional justice. This sensitivity of new RETs' regional benefits and burdens to distributive justice definitions illustrates that equity not only depends on the location of RET infrastructures but also on what different groups and communities perceive as just, e.g., who owns energy infrastructure, wellbeing, job creation, and local environment degradation [42]. Given that Norwegian youth exhibit more favourable attitudes towards RETs than older generations [43], involving young people's representatives in decision-making processes may help foster local acceptance of new solutions to conflicts associated with RETs. As young people residing in coastal regions may not necessarily share such preferences as those in the sampled regions, we suggest combining the national focus of the current study with local considerations through regional case studies to ensure that offshore solutions are truly equitable.

We demonstrate how effective communication, empowerment, and collaborative decision-making with underrepresented groups can foster procedural justice and inclusive solutions, as differences in stakeholder opinions may not be techno-economically difficult to reconcile. A mixture of avoiding and mitigating negative impacts, embedding new RET installations within place-based value creation, and compensating for unavoidable impacts in a desired form can foster distributional justice. A feedback discussion session highlighted that climate actions must be approached as localised social issues (community-based) given Norway's diverse cultural landscape, where a one-size-fits-all approach may not lead to effective policy implementation at local levels. This session also illustrated that co-designed modelling studies enable participants to reflect on choices and assumptions. For example, students reconsidered some onshore wind restrictions after the modelling results revealed that these led to higher system costs. By meaningfully engaging diverse groups, such as youth, in decision-making processes, Norway can better navigate the challenges associated with the energy transition.

Although the scenarios can be tailored based on local socio-political context, the framework employed in this study is applicable to other underrepresented groups and countries, especially in the global north, where discussions about energy transition are happening at wider political discourse due to increased negative public sentiments [16]. Using the proposed framework, local governments can develop guidelines incorporating diverse groups' perspectives within the community, ensuring that new renewable installations align with national and regional plans, and can actually be delivered on-time as opposed to being either blocked or delayed; thus resulting in the energy transition being more costly. This strategic approach can help enable countries to effectively navigate the socio-economic challenges associated with transitioning toward sustainable energy systems.

# 8. Methods

## 8.1 Modelling framework: adaptation of highRES electricity system model

In this Norway-focused study, we modify the open-source high temporal and spatial resolution Electricity System model (highRES), designed to assess high shares of variable renewables and explore flexibility options [32]. Utilizing perfect foresight of input data, highRES optimizes the spatial allocation of new RET installations, operational decisions, transmission infrastructure, and storage technologies while minimizing the total system costs, encompassing annualized investments and operating costs. In evaluating the costs associated with students' preferences, their choices were incorporated into the model through the constrained optimization approach explained below. The hybrid greenfield model does not account for existing solar- or wind-based installed infrastructure, except for hydropower, pumped hydro storage, and transmission infrastructure, including interconnections with other countries. Hydropower capacities remain fixed as they are expected to sustain until 2050, and their alteration could impact ecosystems. We focused exclusively on battery-based energy storage, operating under the assumption that local battery production will gradually increase by 2050 [37].

We exclude investment costs for the existing infrastructure but include fixed and variable operation and upkeep costs. The existing 5 GW of installed wind power capacity is excluded, assuming that these wind farms will be entirely replaced by 2050 due to their lifespan limits, allowing the model to determine whether new wind installations should occur at these locations. We adapt the highRES modelling framework for Norway by incorporating changes to model scenarios, reflecting the youth's preferences across different power system aspects. The modified version, highRES-Norway, and replication data are openly available.

The baseline scenario includes the highRES's implicit settings. We maintained Norway's interconnection capacity at its current level, as reported by Statnett, the Norwegian Transmission System Operator (TSO), with no projections for 2050 [44]. In the base scenario, interconnection capacities are assumed to be utilised at one quarter of their rated capacity based on historical utilization rates [45]. Moreover, the base case represents the current regional transmission capacities, established as a lower bound to enable assessment of grid expansion impacts in line with students' choices. Transmission capacities between Norwegian counties as well as with other countries are given in the **Supplemental Tables 3 and 4.**

For analysis consistency, a fixed import and export price is assumed, based on 2020 hourly electricity prices from connected countries, weighted by the interconnector size. While Norway's administrative regions (called zones in this study) were restructured in 2024, we use pre-structuring boundaries of 11 regions due to the project's initial work and educational workshop preparations. We apply a system-wide carbon emission limit of 0 $gCO_2$/kWh across all demand years, precluding carbon-emitting technologies, including carbon capture and storage. The core modelling framework is employed with the workflow management system Snakemake, automating input data processing based on student choices across power system dimensions and generating all possible sequencing orders for integrating these preferences. The detailed core model structure is documented in Refs [32,35,46,47].

We use 2010 weather data input due to its challenging conditions—unusual cold and reduced precipitation—leading to high electricity demand and less hydropower production, Norway's primary electricity source [48]. ERA5 weather reanalysis climate data from ECMWF [49] are converted into time series for power system variables (i.e., capacity factors) using the xarray-based Python library Atlite [50]. Due to Norway's complex terrain, ERA5's 0.25 degree data insufficiently captures localized wind variations, possibly leading to wind power estimation inaccuracies. Therefore, a bias correction function is computed using historical wind power production data from the Norwegian Water Resources and Energy Directorate (NVE), comparing NVE historical production with ERA5 wind speeds at operational wind parks for each 30 x 30 km grid cell such that the year-round distribution of capacity factors of both datasets matches [35]. This corrected data undergoes further exclusions in the workflow process before being put into the model.

For the investment decisions regarding new RET installations, the spatial level is set at the zonal level, with 30 x 30 km grid cell weather data averaged at this level, excluding poorly performing cells with average capacity factors below thresholds—0.09 for solar, 0.15 onshore wind, and 0.20 offshore wind power [38]. Further exclusions are based on the World Database of Protected Areas, terrain slope over 15 degrees, elevation above 2000 metres, and physical infrastructures (military areas, airports, glaciers, etc.). Additional exclusions are also applied based on buffers around urban fabric (2 km) and other infrastructures (5 km) for safety reasons, per Refs [35]. These exclusions delineate the RET deployment area available in the base scenario. Further exclusions in the base scenario grid cells are based on student landscape preferences, significantly altering the available input area**.** All exclusions are measured in area (km$^2$) and converted to solar and wind capacity potentials at zonal and national levels using the assumed conversion metric of 3 MW/km$^2$ [40].

Like the weather year, the 2010 electricity demand time series serves as a baseline, augmented with future projections for 2030, 2040, and 2050. According to Norwegian TSO forecasts, the maximum Norwegian electricity demand in 2050 is projected at 260 TWh, up from 140 TWh in 2022, including demand from electric transport [37]. The increased demand is distributed spatially across zones at hourly temporal resolution using methods detailed in Refs [47] **(Supplementary Fig. 15)**. The distribution of demand increases across different Norwegian sectors is detailed in Ref [47]. The technology cost parameters used in this study are depicted in **Supplementary Table 2**.

Several assumptions underlying this study could impact the results. The exclusive focus on wind- and solar-based technologies reflects Norwegian governmental priorities without considering existing solar and wind power capacities in light of the 2050 net-zero projections. We extend electricity transmission capacities beyond current levels while respecting students' choices, ensuring alignment with existing infrastructure constraints. Interconnections with new countries were not envisaged; enhancements to current interconnections were emphasized, with import/export costs assumed constant. Additional electricity system modelling assumptions are detailed in the **Supplementary note 1**.

## 8.2 Workshops-driven scenario framework

The methodological details of workshops are given in **Supplementary note 2**. The educational workshops spanned a full day, encompassing various interactive activities specifically designed to foster meaningful participant engagement. By restructuring the modeling approach, we meaningfully engaged young people's perspectives across the entire ESM process. This inter- and multidisciplinary study involved psychologists, political scientists, sociologists, biologists, RET scientists, and energy system modellers, ensuring workshops facilitated collective learning and provided context for informed energy choices.

The questionnaire was aligned with the energy system planning model and addressed several power system aspects, including choices among different RETs and between energy storage and electricity trading (import/export), transmission infrastructure expansion, various techno-economic choices, and preferences regarding regional and landscape settings for new RETs. These core aspects are crucial for capturing energy justice considerations associated with evolving power systems [27]. Following a detailed assessment of workshop data [51], six main scenarios were developed to evaluate the impact of students' preferences on the snapshots of a net-zero electricity system design for Norway. A range of scenarios was simulated based on prioritization and incremental integration of students' choices alongside stepwise exclusions of disagreed landscapes. **Supplementary Table 5** illustrates exemplary representative scenarios reflecting young people's viewpoints.

The baseline scenario is a model-based optimization benchmark for evaluating the student choices' impact. It includes default exclusions discussed in the previous section and maintains current electricity transmission and import/export capacities. The transmission (trans) scenario reflects student preferences regarding whether: 1) to extend the grid, and 2) to choose between expansion with overhead or underground power lines. To incorporate students' choices, current zonal transmission capacities were set as lower bounds, and a transmission equation was added to ensure new capacities encompass the desired percentage of overhead and underground cables:

$$\sum_{z,z'} subsurface \geq \alpha \, . \left[ \sum_{z,z'} pcap - \sum_{z,z'} cap0 \right]$$

Given the model's prioritisation of least-cost overhead investments, the equation ensures proportions of subsurface power lines between zones (z) in the transmission expansion are maintained by subtracting existing capacities (cap0) from total capacities (pcap). Alpha denotes the percentage of participants' support for underground power lines, ensuring that at least this proportion is embodied within the net-zero power system.

The trade scenario reflects young people's choices regarding electricity trading with other countries to address the challenges of renewable variability. Regardless of import/export levels, students favored self-sufficiency—equating imports with exports—and local energy storage. To reflect participants' desires for import/export in trade scenarios, the upper bounds of interconnection utilisation rates were varied up to threefold, assuming full utilisation without new power line investment costs. To reflect the principle of self-sufficiency, the electricity balance equations were redefined to ensure that annual import and export levels matched, thereby enabling more efficient optimization of power dispatch.

$$\sum_{tech} G + \sum_{storage} (out - in) + \sum_{trans} (in - out) = Demand_h + Export_h \text{ subject to: } \sum_h Export_h = \sum_h Import_h$$

Here, import is assumed to be part of generation technologies (G), and export is added to demand. By maintaining self-sufficiency, total annual electricity trade levels were incrementally increased to threefold to evaluate the impact, reflecting participants' support for helping other countries in their energy transition.

Transmission and trade scenarios were further simulated across three transmission (underground, overhead, and fixed transmission lines) and trade (the current level limits, two and three times of the current levels) categories, as the majority of students' selections fell into these categories **(Supplementary Figs 8–10)** [51]. For the primary analysis, scenarios involving underground transmission and electricity trade at twice the current levels were selected, based on their prevalence among students and their representation of moderately ambitious energy transition strategies.

The land scenario conveys students' preferences for landscapes favored or opposed to RET installations. The assessment indicated a strong preference for installations in sea, mountain, and vegetated areas, while placement in forests, agricultural lands, and residential areas was strongly opposed. Landscapes receiving a median value of less than 5 on a 9-point Likert scale were interpreted as reflecting significant opposition to installing new RETs and were therefore excluded from land scenarios for future RET deployment [3,40]. The high-resolution Norwegian land data, from the Norwegian Institute of Bioeconomy Research, and the CORINE data were utilized to spatially exclude grid cells, based on landscape preferences atop the base scenario [52,53]. These exclusions, detailed earlier, lead to substantial changes in available areas for new capacity installations and available generation capacity potential per technology.

The technology (tech) scenario reflects the students' technology choices for achieving a net-zero electricity system. The well-known discrete choice approach was employed to understand the youth's preferences regarding RETs [51]. Based on the quantitative assessment, preference levels for each technology were translated into aggregated upper limits on new capacity installations within the national RET capacity mix. The ratio between fixed and floating installations was optimized for offshore wind since it was difficult for students to differentiate between them.

$$pcap_i \leq \beta_i \sum_j pcap_j$$

Where pcap represents the new capacity of technology type *i*, beta is the preference coefficient derived from participants' preferences through discrete choice analysis, and j represents the set of RET technologies. Preference coefficients reflect the percentage of participants who choose the particular RET, ensuring that its proportion in the new RET capacity mix does not exceed the specified limit. For instance, if the model includes onshore wind, its proportion in the new capacity mix must not exceed 23.5%.

The regional (region) scenario interprets students' choices about appropriate wind and solar farm installation zones. The prior method was used to translate preferences into upper limits on new capacity installations per zone and technology. For instance, preference levels for solar (~13%) and onshore wind (~10%) in Oslo delimit new capacity installations not exceeding these values relative to the aggregated national installations. Accordingly, depending on the capacity potential for each technology, the model may select installations in Oslo, constrained by participants' preferences.

$$pcap_{i,r} \leq \gamma_{i,r}.pcap_i$$

Where pcap signifies the new capacity of technology i in region r, while gamma represents the technology's preference coefficient for a given region derived from students' choices, for example, if the model chooses onshore wind capacity in Oslo, it must not exceed 10%. This scenario is notably conservative, and the model struggles to achieve optimization due to tight zonal constraints. Despite workshop materials and questionnaires being tailored to highRES model characteristics to minimise assumptions while translating students' outputs, we acknowledge potential bias arising from interpreting and analyzing the workshop results.

### 8.3 Equity evaluation

Research indicates that equity within energy systems considerably varies according to how the burden or benefits of energy infrastructures are distributed [38]. Beyond establishing an appropriate definition, another crucial aspect involves deciding what should be distributed equitably, such as allocating new technology capacity installations, emission rights, or job opportunities. We adopted three principles that are most pertinent for analysis at the national scale. For land-burden and equality principles, the electricity system's installed capacities (both new and existing) serve as the equity factor, weighted against zone total areas and population. The consideration of existing electricity infrastructure is vital in Norway's context, particularly since hydropower is concentrated primarily in specific zones. For the self-sufficiency principle, total electricity generation is considered an equity factor, weighted by yearly electricity demand, as electricity generation takes precedence over installed capacity for this principle. **Supplementary Fig. 15** illustrates the spatial distribution of denominators used for the justice definitions considered. The study utilises the latest available descriptive statistics for population and areas from [45].

To evaluate the distribution of burdens (i.e., RET capacities and electricity production) across zones and the impact of student choices on equity, burdens are plotted in an ordered weighted format on a Lorenz curve, where each point represents a zone. The divergence between the straight line and the Lorenz curve provides insight into the equity of a given preference scenario, facilitating the calculation of Gini coefficients. This metric frequently features in energy system analyses to gauge inequalities, specifically in assessing how evenly the burden of new RET installations is distributed [54]. The Gini coefficient is "the mean difference from all observed quantities". The detailed mathematical interpretation of the Gini coefficient can be found in Ref. [38]. Having calculated the Gini index, one can determine the sensitivity of the spatial allocation of RET capacities to the regional equity. It is worth mentioning that the relationship between Gini variables is not linear and, hence, cannot be incorporated into a linear optimisation model. Instead, Gini values are calculated post-processing to consider the different framings of distributional justice.

### 8.4 Limitations and future work

While our results demonstrate the feasibility and potential of using school workshop-informed modelling to elicit underrepresented groups' participation in designing the future net-zero energy system, we suggest extending this work in two ways: by increasing representation and by conducting follow-up sessions. A next step to ensure broader youth representation involves expanding the geographic range to cover central and Northern counties, as well as relatedly, different segments, such as coastal, mountain, or Sami communities. Comparing the reasoning and the resulting scenarios, preferably by organizing discussion rounds across the country, is expected to help researchers and young people better understand the different perspectives of the energy transformation. Such an approach highlights the value of youth-informed scenarios as a tool for fostering future-oriented thinking and inclusive participation.

Yet, such models' apparent precision and accuracy can obscure the inherent uncertainties embedded in underlying assumptions. Both the techno-economic and socially-determined parameters are influenced by normal fluctuations and larger events like pandemics, wars, tariffs, extreme climate events, terrorism, sharp political shifts nationally or abroad, technological failures, unforeseen impacts, and breakthroughs. Follow-up engagement processes will therefore be essential to understanding how such unexpected and large-scale events shape preferences and scenarios' outcomes.

## Supplemental file

Additional results, details of techno-economic modelling input data and conducted workshops are available in the supplemental file.

### Acknowledgements

This study is part of the Energy for Future project, funded by UiO: Energy & Environment. The funding institution was not involved in conducting the workshops. We thank all the schools for support in organizing and conducting workshops and all students for their contributions.

We would like to thank Oskar Vågerö for input through discussions during and after the workshops.

### Data and code availability

The modified version of highRES electricity model used in this paper is open source and available at **Github** (https://github.com/JavedMS/highRES-Norway.git), containing the model formulation, data necessary to replicate the simulations and analysis as well as the modified model documentation describing the changes from previous versions. The detailed workshop methodology and modelling questionnaire used are available in Ref [51]. The materials of workshop and interactive activities can be provided upon request. High-resolution Norwegian land data used for land exclusions is openly available at [52].

# Supplemental file

# Engaging young people for a more inclusive national energy transition: A participatory modelling framework

Muhammad Shahzad Javed, Karin Fossheim, Maximilian Roithner, Matylda N. Guzik,

James Price, Beate Seibt, Marianne Zeyringer

**Supplementary Table 1** presents a selection of studies that linked social theories with the existing energy system models. While not exhaustive, this table illustrates the increasing engagement of participatory modelling, including recommendations and challenges inherent in this approach. We only present studies involving physical interactions with stakeholders or direct data collection through stakeholder engagement, excluding those relying on online public surveys or existing social science literature data. The "level of integration" describes how the stakeholders' inputs are incorporated into energy system modelling (ESM). A "shallow" level indicates that social considerations are added as a side/separate work to the techno-economic analysis of ESM, whereas a "meaningful" level represents equal weightage being given to societal and techno-economic dimensions, or at least collecting stakeholders' input before projecting energy scenarios or designing the research framework. Though feedback sessions help participants reflect on the impact of their perspectives and choices on the future net-zero energy systems, to the best of the authors' knowledge, no study has conducted a formal feedback session with participants, and very few studies involve more than one interaction with them.

**Supplementary Table 1.** Outcomes/recommendations of selected participatory studies. The level of integration is defined based on McGookin et al. [1,2].

| Study | Level of integration * | Reference | Outcomes/recommendations |
|---|---|---|---|
| McGookin et al. (2022) | Meaningful | [3] | • Clear tension between community revival and emission reduction ambitions.<br>• More collaboration with diverse stakeholders is needed throughout the policy process. |
| Li FGN et al. (2016) | Meaningful | [4] | • Contradiction between proposed sub-national and national-level optimal transition pathways. |
| | | | ● Without sub-national economic development, equitable energy transition can not be achieved. |
| Fortes et al. (2014) | Meaningful | [5] | ● Policies need to favor the elements that stakeholders respect.<br>● Collaborative future scenario building is necessary to establish the collective capacity of “thinking for future”. |
| Mathy et al. (2016) | Meaningful | [6] | ● Stakeholder scenarios cut emissions up to 80%, benefiting economic growth.<br>● Co-development processes may gain the acceptability of low-carbon trajectories. |
| Venturini et al. (2019) | Meaningful | [7] | ● The gap between energy model projections and real evolution is larger with stakeholders’ input into the energy model.<br>● Conveying the model's structure may help to sensibly capture social aspects in the modeling framework. |
| Venturini et al. (2019) | Meaningful | [8] | ● Participatory scenario-making leads to a shared understanding of future energy system pathways. |
| Bertsch et al. (2016) | Meaningful | [9] | ● Paradoxical discrepancy between low public acceptance of grid expansion and high acceptance of environmental sustainability and security of supply. |
| Zelt et al. (2019) | Meaningful | [10] | ● Regardless of country background, participants preferred a high share of renewables for future energy systems.<br>● Revised policies with participatory process are recommended. |
| Chapman et al. (2018) | Meaningful | [11] | ● Favoring a socially equitable energy system positively affects the transition to low-carbon energy systems.<br>● Among many people's choices, electricity price and employment greatly impact energy policy. |
| EKER et al. (2018) | Meaningful | [12] | ● Monitoring and energy efficiency considerations improved residents' well-being, suggesting that awareness and management of energy usage positively affect residents' perceptions and comfort.<br>● Participatory approach-based decision-making is recommended for policymaking. |
| McKenna et al. (2018) | Meaningful | [13] | ● The stakeholders do not pursue the maximization of the economic sustainability scenario.<br>● Robust and concrete recommendations are derivable based on stakeholders’ input. |
| Heaslip et al. (2018) | Meaningful | [14] | ● The iterative process revealed that participants felt heard and owned the inclusive key decisions. |
| Drechsler et al. (2017) | Meaningful | [15] | ● Minor compromises on technology's techno-economic metrics lead to a significant increase in equity.<br>● Equitable allocation of wind/solar calls for a trade-off between efficiency and equity. |
| Fell et al. (2020) | Shallow | [16] | ● Recommended changes in the energy model to represent social factors.<br>● Ensuring transparency by breaking down policy costs on energy bills reduces the risk of inequitable outcomes in the low-carbon transition. |
| Koecklin et al. (2021) | Meaningful | [17] | ● System costs increase up to 33% by considering people's choices.<br>● Energy transition projections ignoring public acceptance are likely to be sub-optimal. |
| McDowall W. (2012) | Shallow | [18] | ● Consumers remain conservative to accommodate new technologies.<br>● Modeling results are sensitive to social factor assumptions. |
| Schinko et al. (2019) | Meaningful | [19] | ● A participatory approach may increase the political feasibility of energy transition.<br>● Recommendations for participatory scenarios are drawn to de-risk investments in energy transition. |
| Sharma et al. (2020) | Meaningful | [20] | ● Stakeholders' understanding of investment costs for various low-carbon technologies improved after participating in engagement processes.<br>● Transdisciplinary research leads to a diversity of perspectives and decreases the risk involved in the energy transition. |
| Kowalski et al (2009) | Shallow | [21] | ● Stakeholders' engagement helps address their uncertainties and makes the process democratic<br>● Codesigning scenarios can better capture the decision-making complexity |
| Trutnevyte et al. (2012) | Meaningful | [22] | ● Linking stakeholders and decision-makers may help sustainable decision-making and promote mutual learning. |
| Grafakos et al. (2015) | Meaningful | [23] | ● Physical interactions among decision-makers and stakeholders help mitigate the inconsistencies in preferences for local clean energy developments. |
| Schmid et al. (2012) | Meaningful | [24] | ● Participatory scenarios reveal that the transformation to low-carbon energy systems requires as much societal effort as it does engineering. |

**Supplementary Table 2.** Technology cost parameters in 2024 Norwegian krone (NOK). The costs are expressed in thousands. Overnight capital cost (CAPEX), Fixed operation & maintenance cost (FOM), Variable operation & maintenance cost (VOM).

| Technology | CAPEX(per MW) | FOM(per MW) | VOM (per MWh) |
|---|---|---|---|
| Solar | 515.26 | 105 | 0 |
| Onshore wind | 1392.09 | 420 | 0.106 |
| HydroRoR | — | 874.88 | 0.042 |
| HydroRES | — | 874.88 | 0.042 |
| Offshore wind bottom fixed | 2603.18 | 1785 | 0.035 |
| Offshore wind floating | 3904.74 | 2677.5 | 0.053 |
| Lithium-ion battery | 216.51 | 123.9 | 0 |
| HVAC overhead line + substation cost | 0.21 (per Km) + 45.36 | — | — |
| HVDC subsurface line + substation cost | 2.05 (per Km) + 93.345 | — | — |

**Supplementary Table 3.** Transmission capacities between Norwegian counties (considered as zones in this study)

| Zone 1 | Zone 2 | Link capacity (MW) |
|---|---|---|
| Oslo | Viken | 3000 |
| Rogaland | Vestfold og Telemark | 900 |
| Rogaland | Agder | 1200 |
| Rogaland | Vestland | 750 |
| Møre og Romsdal | Innlandet | 500 |
| Møre og Romsdal | Vestland | 3000 |
| Møre og Romsdal | Trøndelag | 1350 |
| Nordland | Trøndelag | 1350 |
| Nordland | Troms or Finnmark | 600 |
| Viken | Innlandet | 7000 |
| Viken | Vestfold og Telemark | 500 |
| Viken | Vestland | 3900 |
| Innlandet | Trøndelag | 600 |
| Vestfold og Telemark | Agder | 1200 |

**Supplementary Table 4.** Current Norwegian power transmission links with other countries and capacities

| Connected country | Link type | Link distance (Km) | Link capacity (MW) |
|---|---|---|---|
| Germany | HVDC subsea | 1447 | 1400 |
| Denmark | HVAC400KV | 907 | 1700 |
| United Kingdom | HVDC subsea | 1409 | 2800 |
| Neatherland | HVDC subsea | 1365 | 700 |
| Sweden | HVAC400KV | 275 | 3995 |

**Supplementary Table 5.** Exemplary scenarios demonstrating adjustments in parameters based on pupil choices. All potential scenario orders emerging from pupil choices were simulated. For a comprehensive assessment of the resulting pupil-driven scenarios, refer to Ref. [25].

| | Total technology capacity (percent of new capacity additions) | | | Landscape preference | Location-based technology preference (Total=11 counties) | | Power lines | | Import /export |
|---|---|---|---|---|---|---|---|---|---|
| | Solar | Onshore | Offshore | | Solar | Wind | OH | UG | |
| Base scenario | OPT | OPT | OPT | OPT | OPT | OPT | OPT | OPT | OPT |
| Scenario #01 | 25.8 % | 23.5% | 50.7% | OPT | OPT | OPT | OPT | OPT | No change |
| Scenario #02 | 25.8 % | 23.5% | 50.7% | Exclude disagreed | OPT | OPT | OPT | OPT | No change |
| Scenario #03 | 25.8 % | 23.5% | 50.7% | Preferred landscapes | OPT | OPT | OPT | OPT | No change |
| Scenario #04 | 25.8 % | 23.5% | 50.7% | Preferred landscapes | Pupil choice | Pupil choice | OPT | OPT | No change |
| Scenario #05 | 25.8 % | 23.5% | 50.7% | Preferred landscapes | Pupil choice | Pupil choice | 41% | 55.3% | No change |
| … | … | … | … | … | … | … | … | … | … |

OPT: optimal; OH: overhead; UG: underground; No change: same as today

**Supplementary Table 6.** Overview of students and schools. The workshops were conducted with upper secondary students, aged 15-16, across five schools, comprising 286 participants.

| **School** | **Number of classes** | **Number of participants** | **Workshop format** | **Location** | **Study program** |
|---|---|---|---|---|---|
| 1 | 1 | 21 | Spread out | Urban area | General |
| 2 | 1 | 12 | Thematic day | Rural area | Vocational |
| 3 | 3 | 60 | Thematic day | Rural area | General |
| 4 | 1 | 31 | Thematic day | Urban area | General |
| 5 | 6 | 162 | Thematic day | Rural area | General |

## Supplemental Results

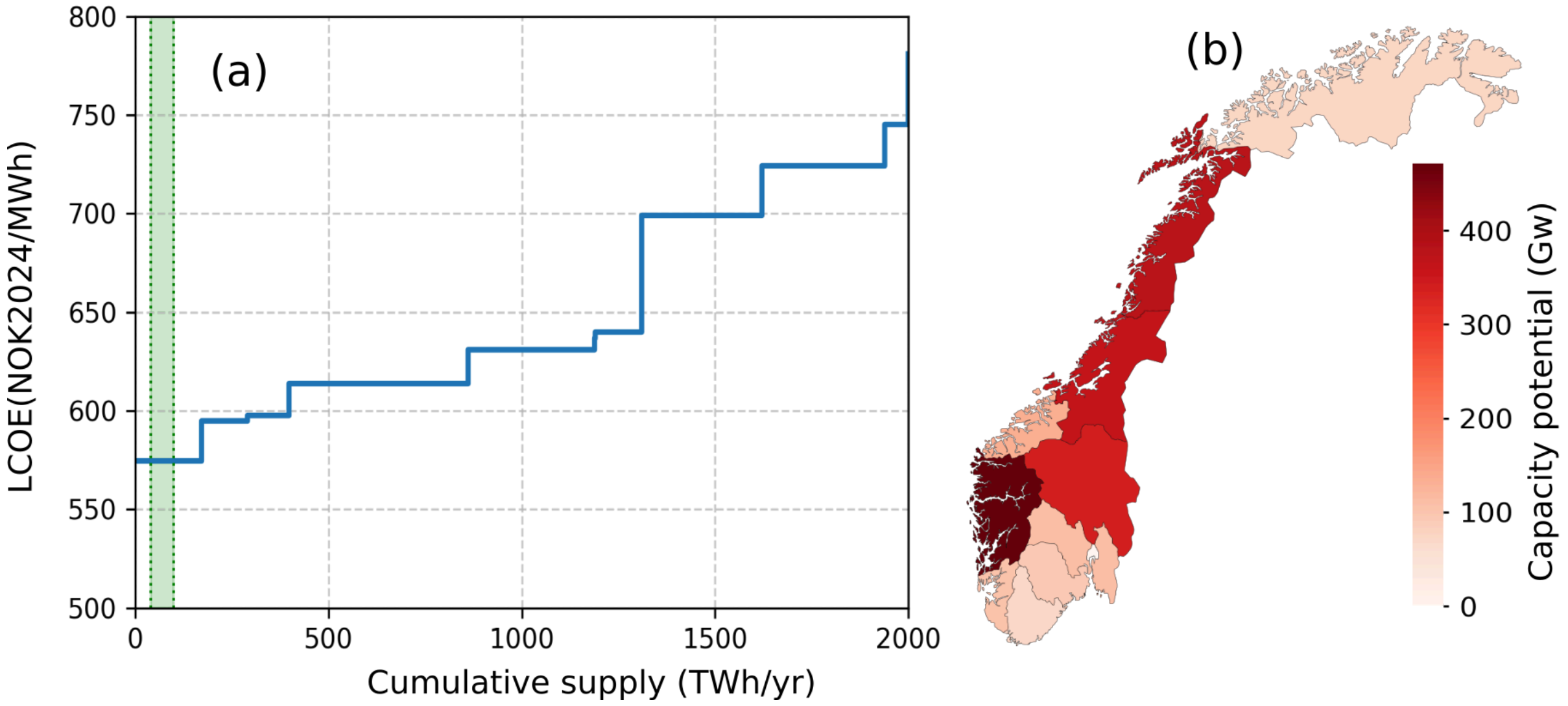

**Supplementary Figure 1:** Cost and capacity potential of solar-based electricity generation in Norway. (a) Cost-potential and supply curve for solar installations, with each step corresponding to a zone. The light green area represents the projected increase in electricity demand by 2050. As of 2024, the grid-connected installed capacity of solar in Norway was 767 MW; (b) Solar electricity generation capacity potential per zone. Available areas are aggregated into zones and subsequently converted to capacity potential. Since student opposed landscape exclusions are applied only to onshore wind, these results reflect the base case [Related to Fig. 2].

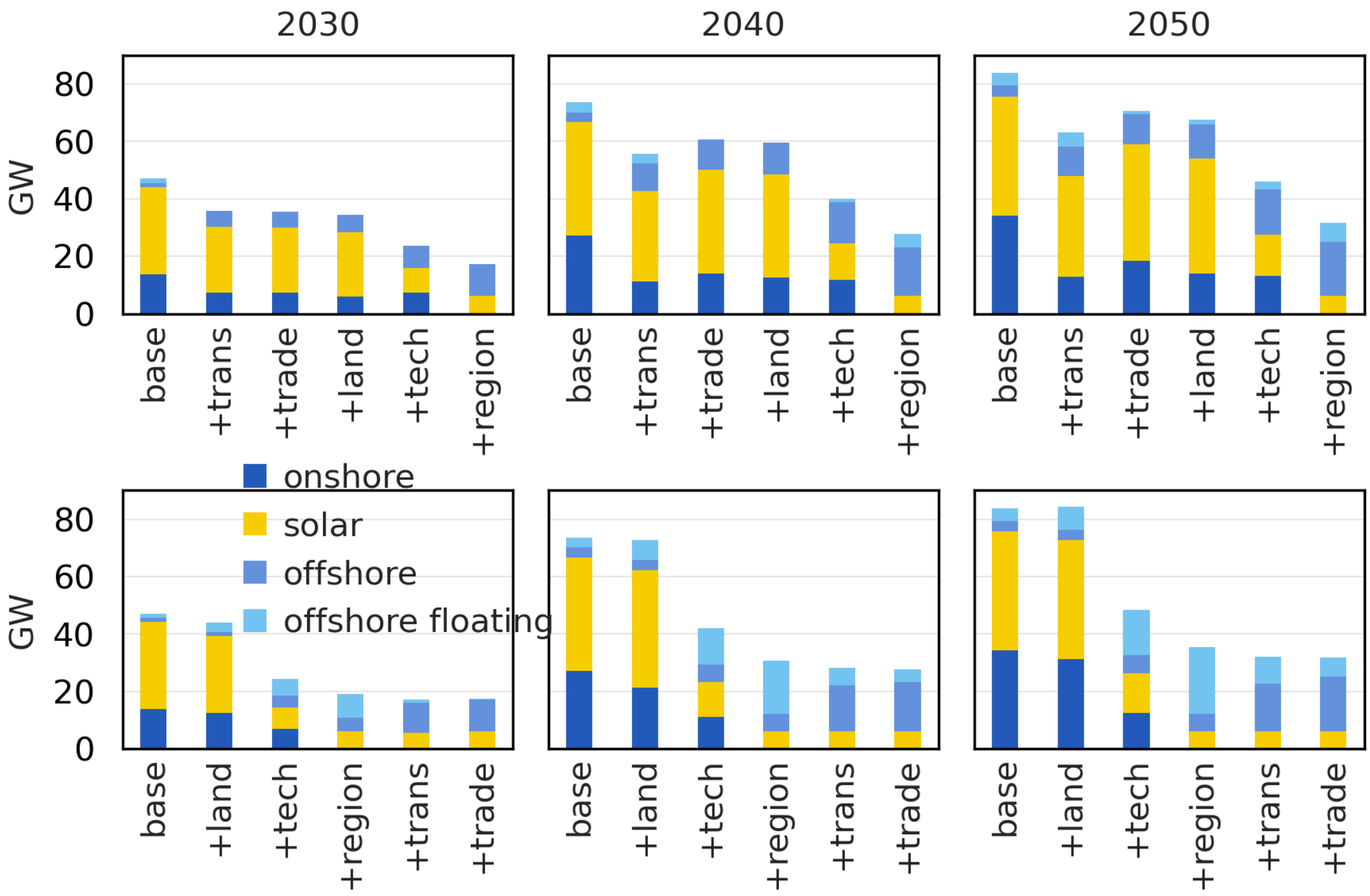

**Supplementary Figure 2:** Snapshots of total new renewable energy technology (RET) installed capacities for the demand years 2030, 2040, and 2050. Regardless of the year, these snapshots demonstrate that the prioritization order of pupil choices substantially influences the cumulative impact. Prioritising trans and trade scenarios always reduces the total RET capacities required to meet electricity demand. Although onshore wind's

role is susceptible to pupil preferences, offshore wind and solar emerge as crucial in achieving an inclusive net-zero energy transition [Related to Fig. 3].

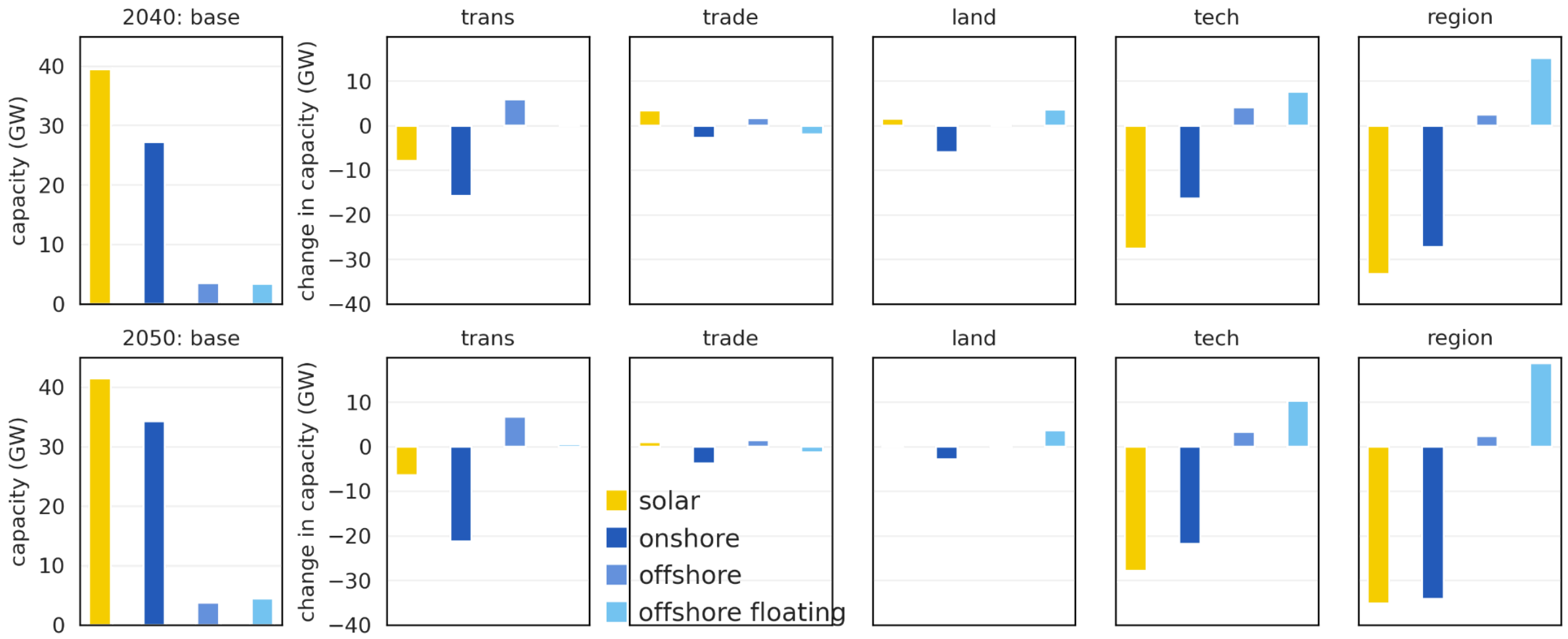

**Supplementary Figure 3:** National installed capacity for the base case and change in RET capacities from the base case when pupil choices are integrated into the model for the demand years 2040 and 2050. Technology and regional preferences substantially reduce the role of solar and onshore wind, leading to increased capacities of offshore wind [Related to Fig. 3].

## Storage capacities:

Although hydropower dominates Norway's electricity system (>85%) and remains an effective resource to mitigate the RET variability, results indicate additional energy storage needs by 2050. Depending on choices made, storage capacities fluctuate between 70 and 125 GWh **(Supplementary Figs. 4–6)**. Increased energy system flexibility, through more transmission and higher trade, reduces storage requirements by ~50% from the baseline scenario (**Supplementary Fig. 4-blue bars)**. Conversely, starting with landscape preferences results in a ~27% increase in the storage capacity, which is subsequently offset by these flexible choices (**Supplementary Fig. 4-brown bars**). Spatial deployment of new storage capacity complements the hydro resources to support new onshore wind in the baseline scenario **(Supplementary Fig. 6, Fig. 6)**. In contrast, regional choices allocate new storage to zones having hydro storage, correlating with new offshore floating capacity (~58%) located near western zones.

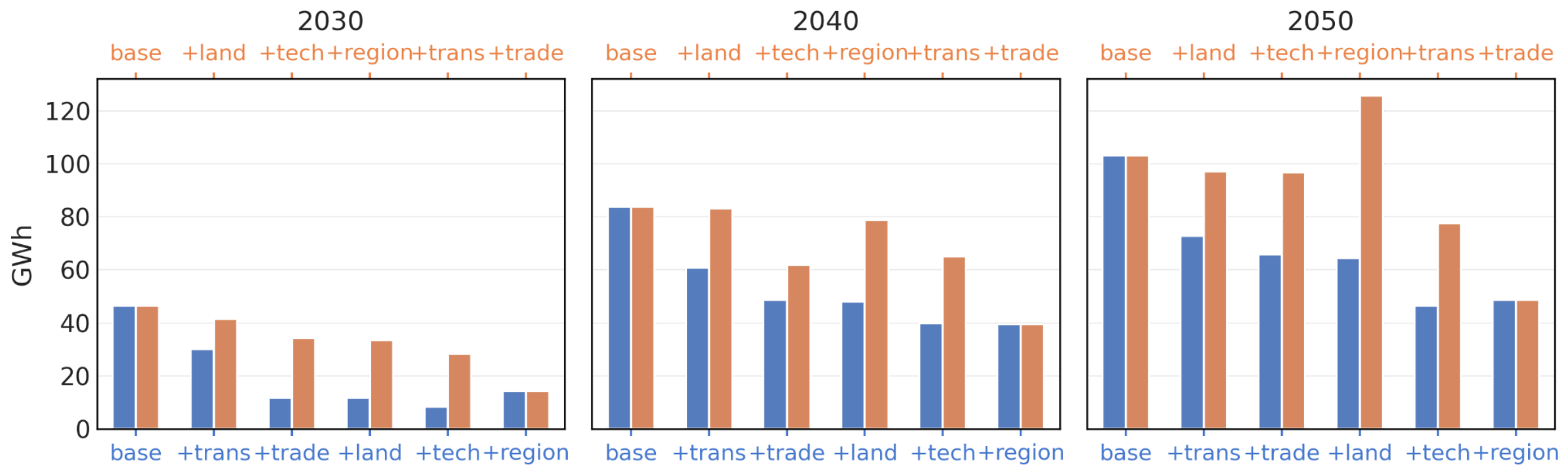

**Supplementary Figure 4:** The effect of incrementally incorporating pupil choices into the base case on new storage capacity requirements for the demand years 2030, 2040, and 2050. The added flexibility of pupil trans and trade choices counteracts the effects of other restrictive scenarios. However, the outcome is contingent upon the prioritisation order of these choices.

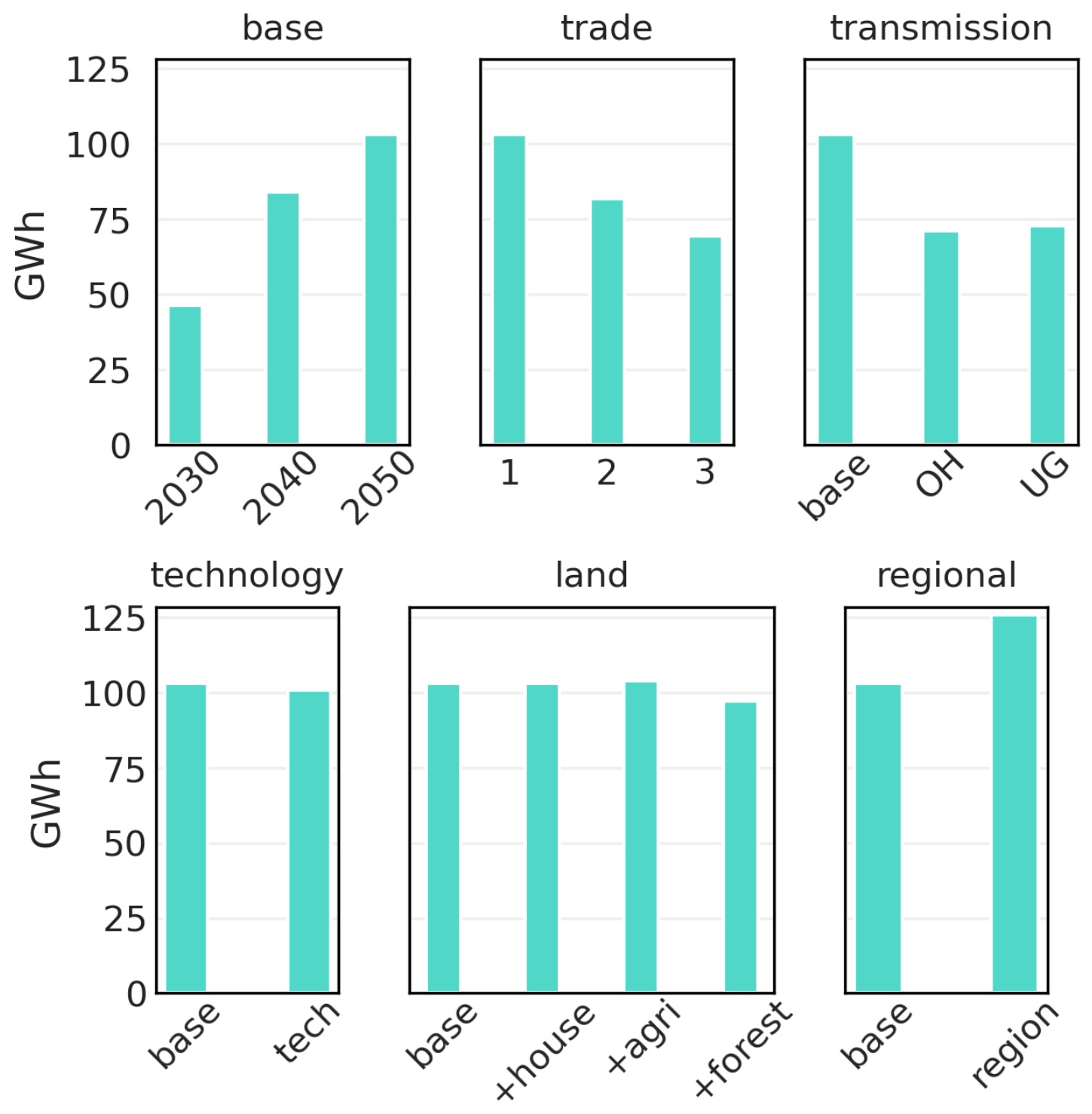

**Supplementary Figure 5:** The individual impact of each pupil choice on storage capacity for the year 2050. The base case reflects the model's optimal storage capacity for the given years. The trade scenario illustrates the impact of increased electricity trading from current levels, while the transmission scenario highlights the impact of overhead (OH) and underground (UG) power line choices on storage capacity. Technology choices and landscape preferences exhibit marginal effects, whereas regional priorities significantly impact storage capacity.

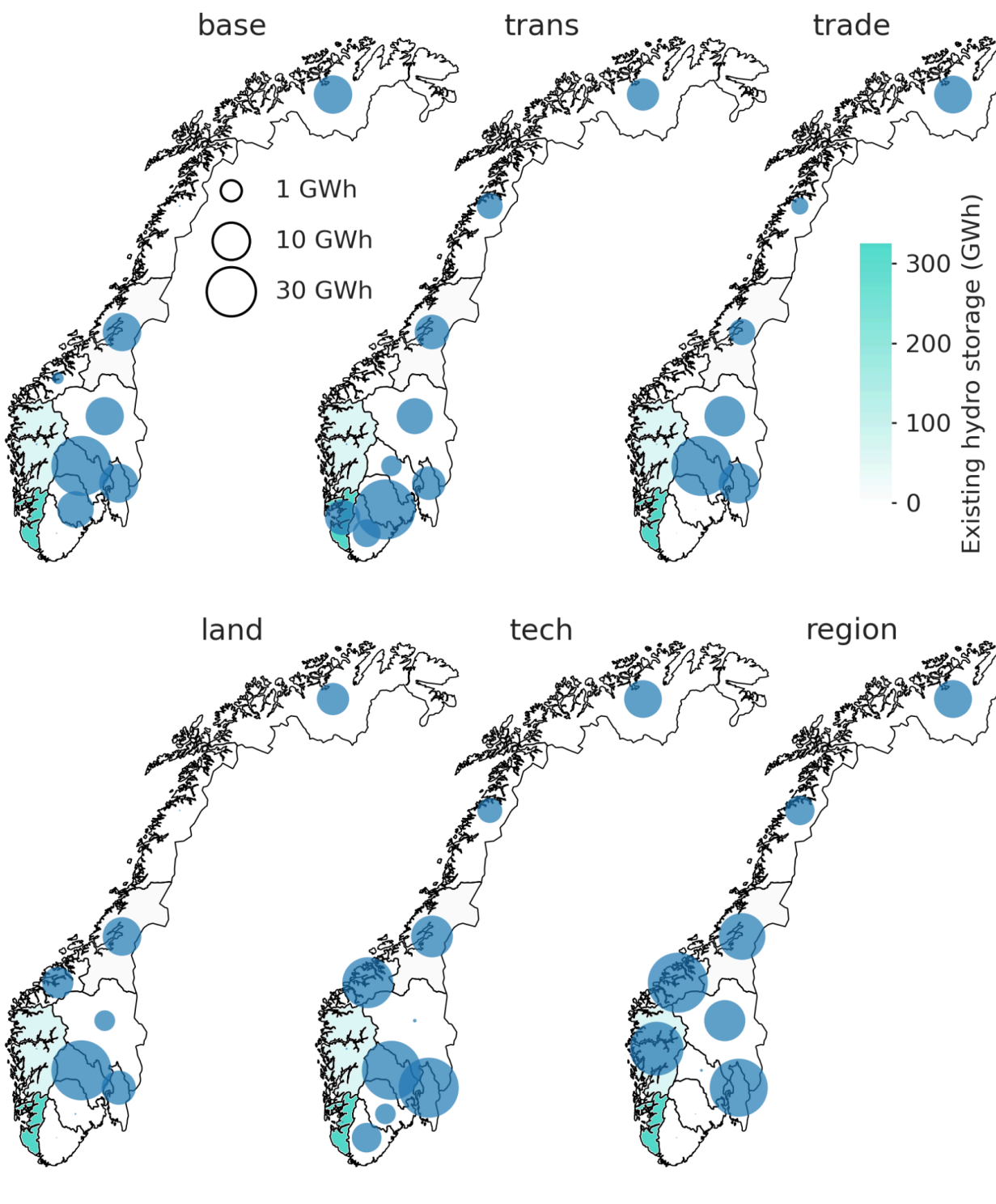

**Supplementary Figure 6:** Spatial distribution of new storage capacity concerning pupil choices for the 2050 net-zero electricity system. Capacity shifts from east to west in land, tech, and region scenarios, largely due to the increased saturation of offshore wind installations in the west. In contrast, trans and trade scenarios allocate new capacities predominantly near demand centres in the east and south.

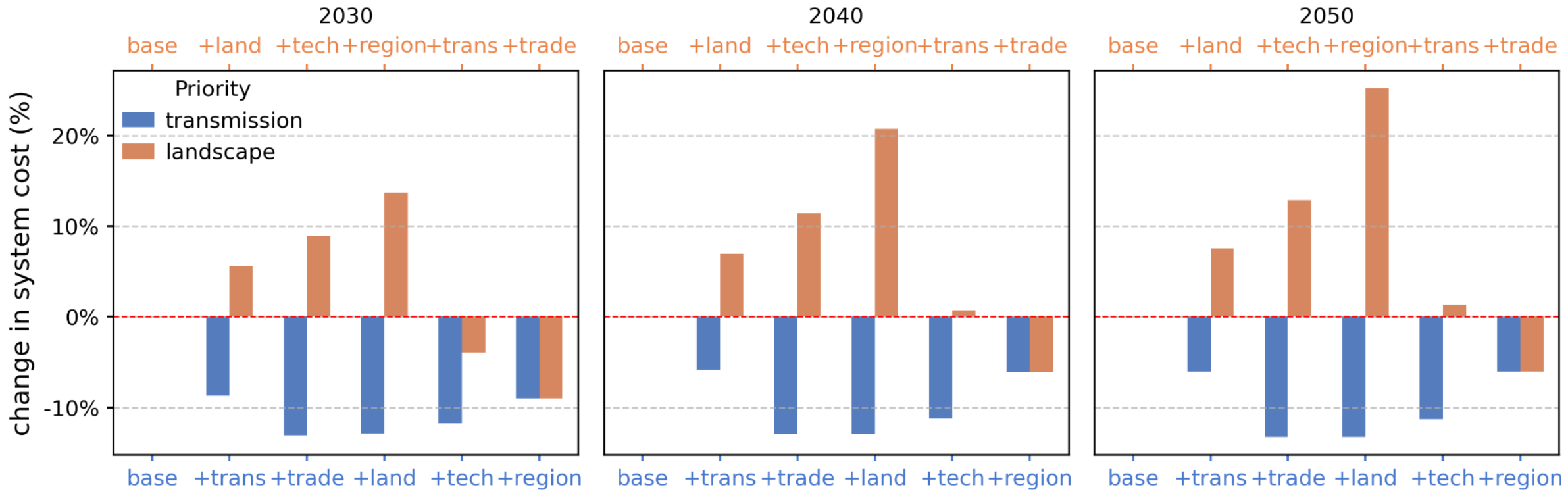

**Supplementary Figure 7:** The cost implications of pupil choices for the specified demand years. Similar to the new RET installed capacities, the cumulative impact on system cost is significantly influenced by the prioritisation order of pupil choices. Irrespective of the demand year, beginning with flexible trans and trade scenarios consistently results in cost reductions of 5% to 10% from the base case [Related to Fig. 5].

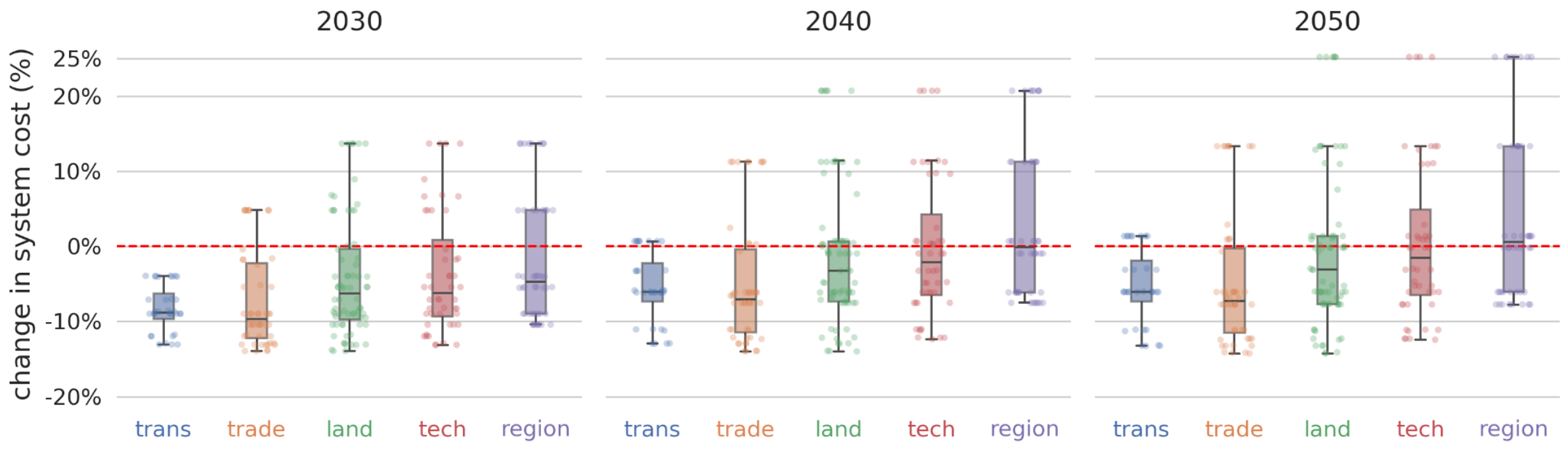

**Supplementary Figure 8:** Boxplots illustrating the sensitivity of system cost across all possible prioritisation orders of pupil choices. These combinations were simulated using the modelling-to-generate-alternatives approach, with the highest priority assigned to specified scenarios. System cost variations range from -12 % to +25%, contingent on the scenario order. The dotted red line indicates the change from the base scenario. [Related to Fig. 5]

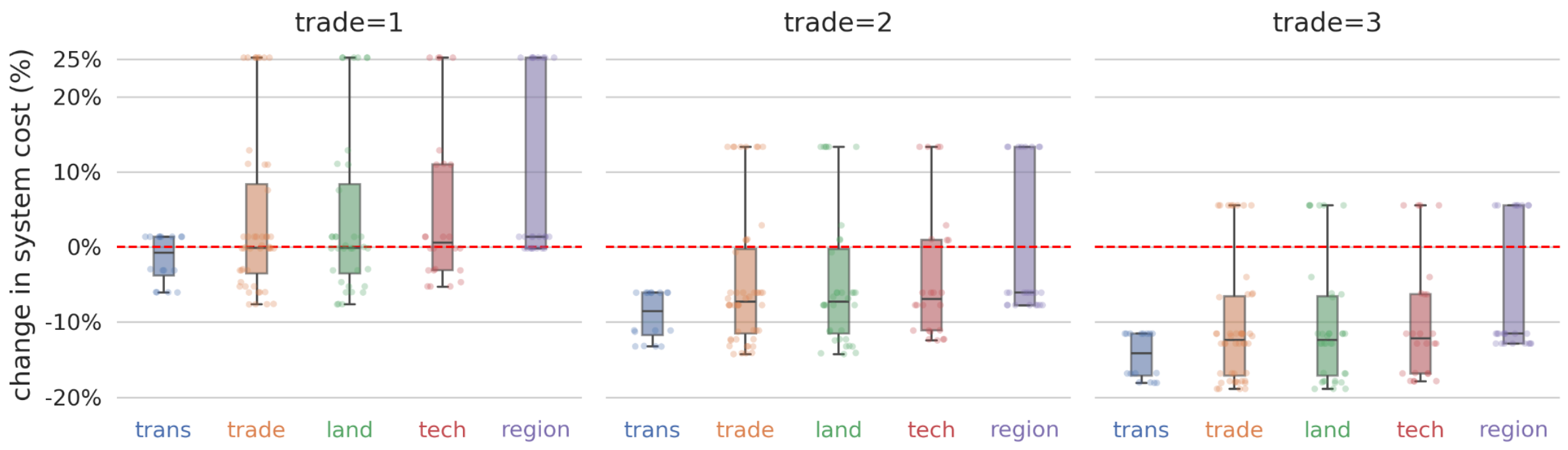

**Supplementary Figure 9:** Boxplots demonstrating the sensitivity of system cost with respect to various trading levels, with trade=1 representing current electricity trading levels. Total annual electricity trading levels increased from the current baseline, whilst maintaining the pupil preference for self-sufficiency— ensuring imports equal exports. These scenarios were simulated in consideration of pupils' support for aiding other countries in mitigating climate change impacts [Related to Fig. 5].

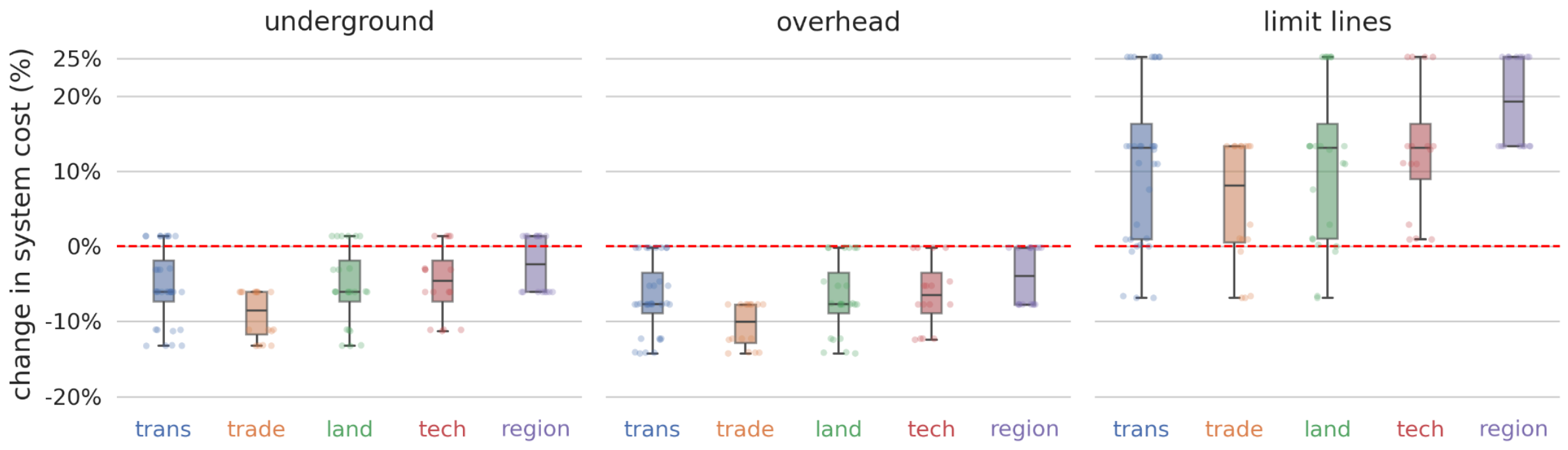

**Supplementary Figure 10:** Boxplots illustrating the sensitivity of system cost across various transmission choices. These scenarios represent pupil preferences for underground and overhead power lines, while limit lines mirror current transmission infrastructure. Regardless of the type of power lines, expanding transmission infrastructure mitigates the cost increase impact from other pupil choices, resulting in median system costs consistently lower than the base case [Related to Fig. 5].

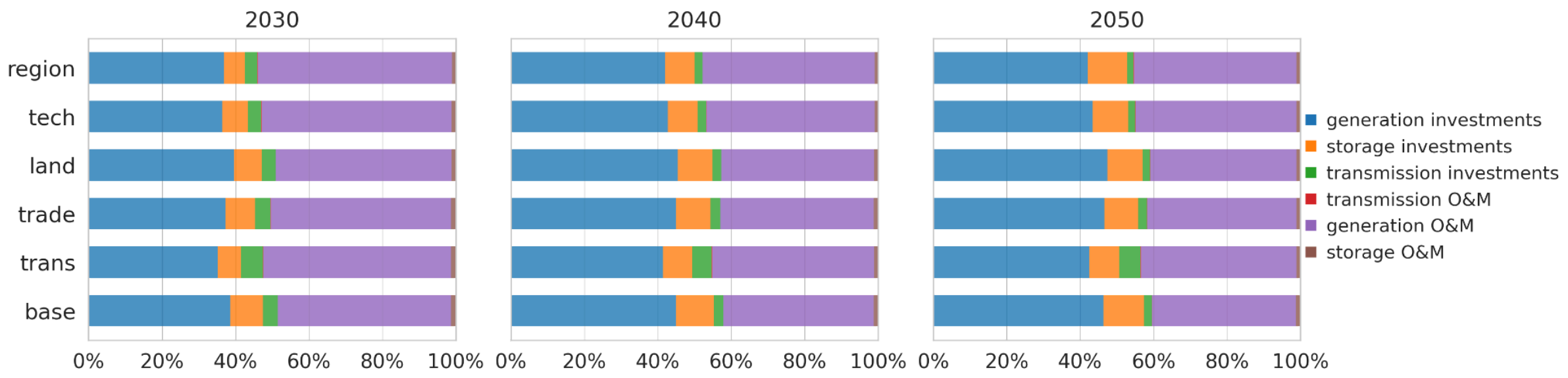

**Supplementary Figure 11:** System cost compositions across scenarios for the demand years 2030, 2040, and 2050. Upkeep and maintenance costs are the predominant components due to the expenses associated with maintaining existing infrastructure and the high maintenance costs of offshore wind installations [Related to Fig. 5].

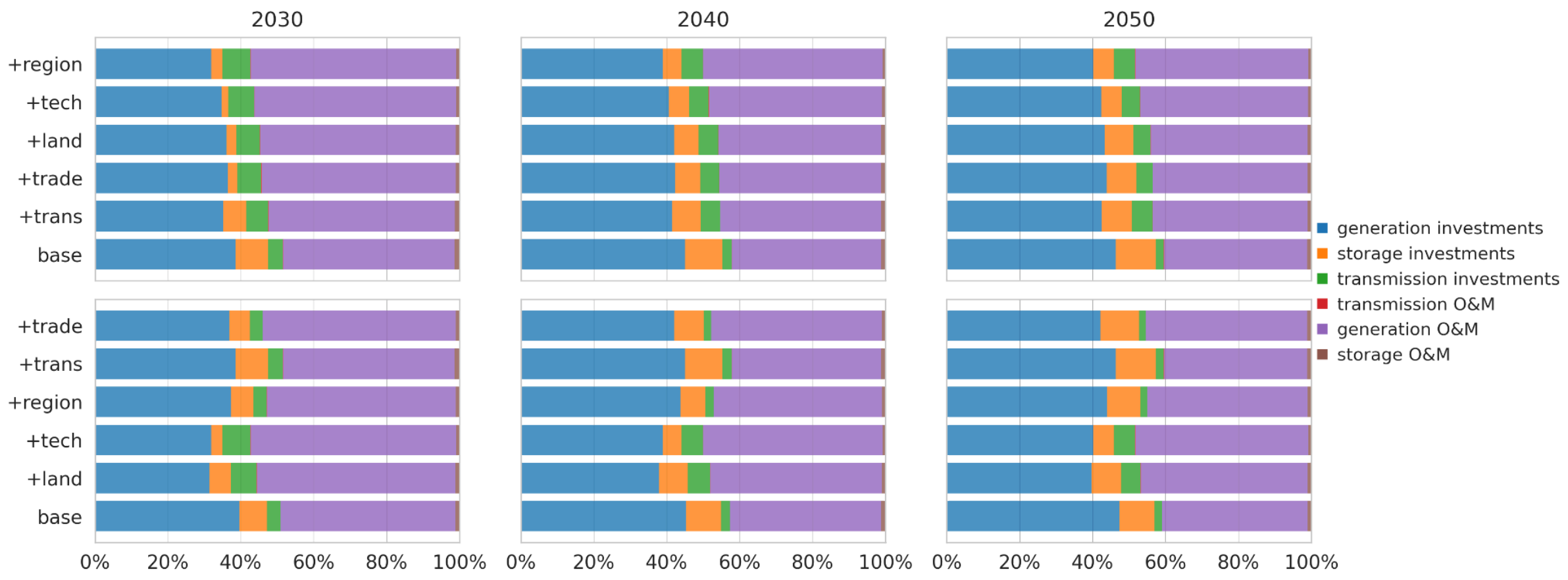

**Supplementary Figure 12:** The cumulative impact of pupil choices on system cost compositions for the demand years 2030, 2040, and 2050. Incrementally integrating pupil choices leads to an increased proportion of operation and maintenance costs [Related to Fig. 5].

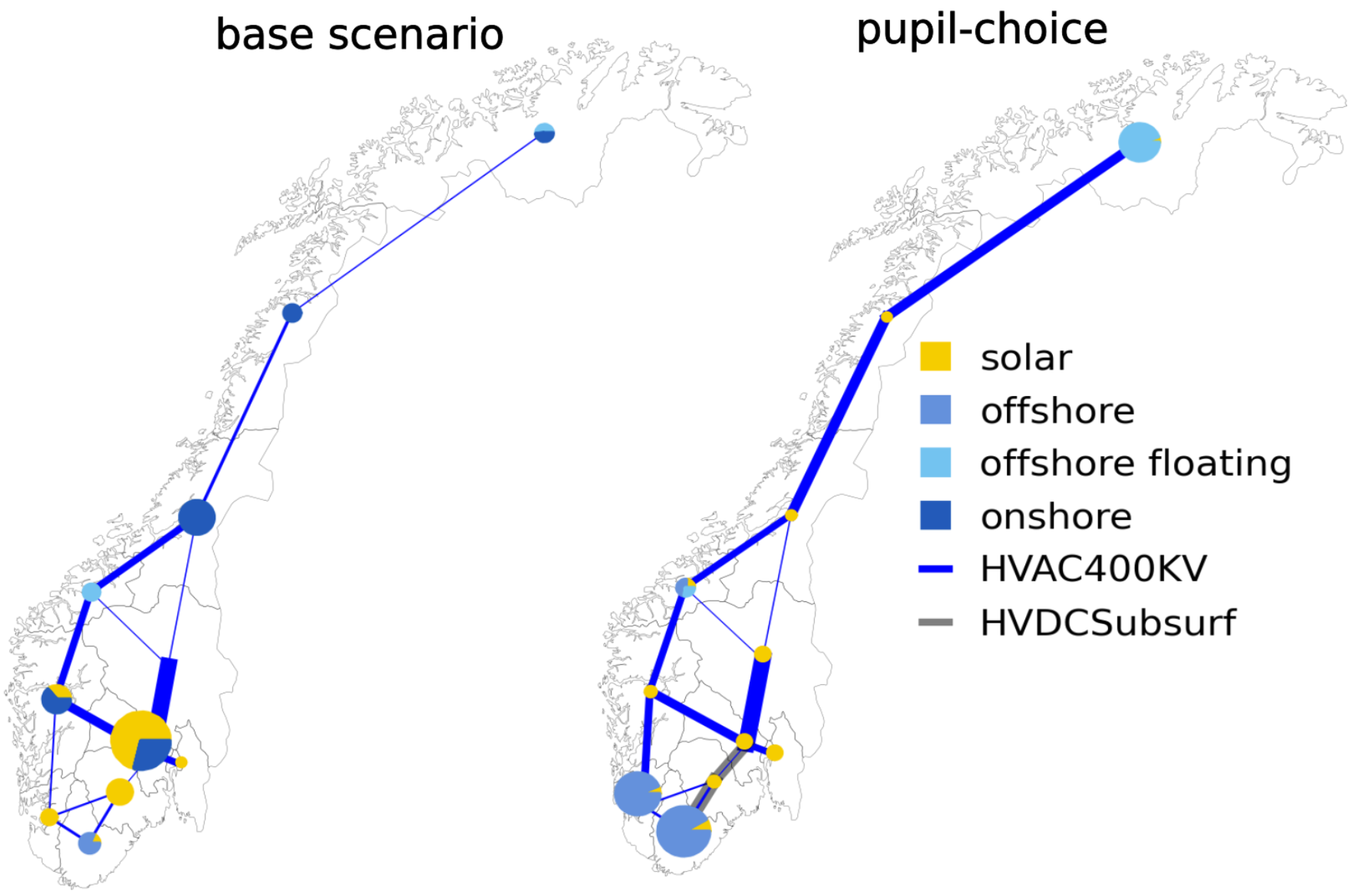

**Supplementary Figure 13:** Electricity system design for 2050, comparing the base case with the pupil-envisioned scenario. The figure illustrates the spatial distribution of new RET and transmission (existing plus new) capacities across Norwegian zones. Pupil choices resulted in more dispersed solar installations, increased saturation of offshore wind, and nearly doubled power transmission lines capacity, including investments in high-cost underground power lines [Related to Fig. 6].

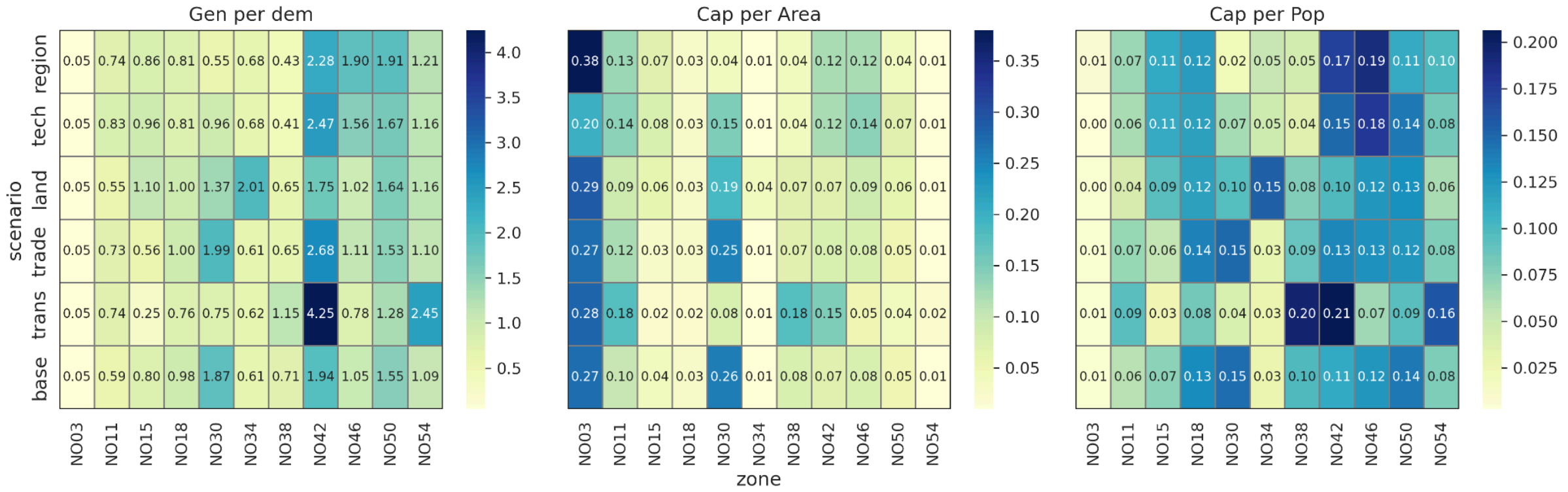

**Supplementary Figure 14:** The relationship between different zones' performance, concerning pupil choice across different power system aspects and applied justice principles: self-sufficiency (gen per dem), land-burden (cap per area), and equality (cap per pop). The results are shown for the 2050 electricity system design. It displays normalized values for "cap per area" and "cap per pop" to enhance understanding. Units for subplots refer to the definitions as gen per dem (GWh/GWh), cap per area (GW/km$^2$), and cap per pop (GW per person). The figure highlights significant changes in the equity impacts of choices based on distributional justice definitions. For instance, all zones outperformed Oslo (NO03) under the capacity per population principle, which performed best under the capacity per area principle due to its relatively small area [Related to Fig. 7].

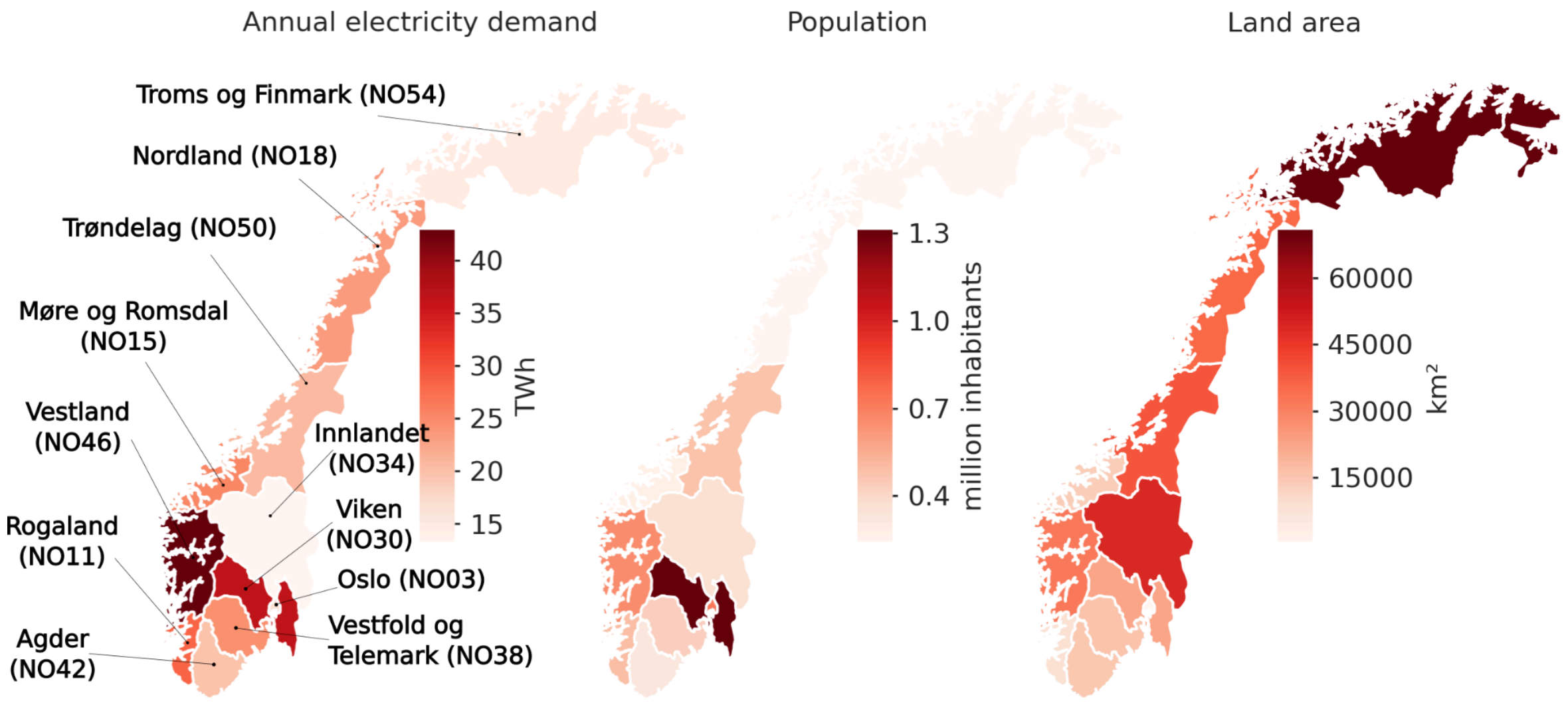

**Supplementary Figure 15:** Spatial plot of equity denominators employed in Gini coefficient calculations, sourced from Norway's Bureau of Statistics [Related to Fig. 7] [26].

## ***Supplementary Note 1***

Our results are based on several assumptions:

- We focused exclusively on solar and wind RET due to the Norwegian government's strong emphasis on these technologies for 2050. Though discussions about nuclear energy are occurring in Norwegian media and among policymakers, these remain speculative and far from concrete plans. Given Norway's ambitious outlook on both offshore bottom-fixed and floating technologies, these were integrated into our modelling analysis. However, during the workshops, we did not elicit separate preferences regarding bottom-fixed vs. floating turbines. Nonetheless, we ensured total offshore wind proportions were based on pupil preferences, with high-resolution models optimising the balance between bottom-fixed and floating offshore wind capacities.

- Translating stakeholders' perspectives into energy system planning has been identified as a key challenge in participatory modeling [2,7]. To minimize translation bias, the questionnaires were designed considering the highRES model limitations, reducing the assumptions needed to integrate pupil choices. As described in the methods sections, we calculated preference coefficients derived from pupil choices, integrating them into the model through various equations. The literature has used this approach to incorporate stakeholders' technology preferences [17,27].

- We acknowledge that pupil reflections on different power system aspects can be influenced by how information is presented and questionnaires are structured. While these biases cannot be entirely eliminated, efforts to minimise them involved an interdisciplinary approach, incorporating social psychologists, political scientists, sociologists, RET scientists, and energy system modellers. For details on workshop content, questionnaire design, and administration, refer to Ref [25].

- We assumed the same lifetimes for both wind turbines and solar panels during workshop demonstrations and modelling, which may slightly underestimate wind turbines' levelised cost of electricity (LCOE) compared to solar photovoltaics due to a slightly lower perceived lifetime (15-20 years). This may impact increased LCOE calculations for satisfying the 2050 electricity demand due to opposed landscape exclusions (Fig. 1). However, given the current difference between the two technologies, this error is minor, with expectations for more significant cost decreases in wind turbines than solar. Additionally, we assumed current transmission infrastructure would remain operational, with slight upgrades by 2050, as these infrastructures' lifetimes are projected to span several decades [28].

- We do not consider the opportunity costs of land when excluding pupil-opposed landscape types, such as agricultural land. While opportunity costs exist, they are relatively minor compared to the investment costs and operational expenses of energy projects.

- Although the workshops were held in the eastern and south-eastern regions of Norway, areas with higher population density than other parts of the country, the projected outcomes based on pupil preferences should not be assumed to represent all of Norway. These results may vary with broader perspectives from across the country and could be the subject of future research.

### *Supplementary Note 2*

Educational workshops were organized and conducted in Norway's eastern and south-eastern regions during the winter and spring of 2024. A total of 286 upper secondary school students, aged 15-16, participated in workshops across five schools (**Supplementary Table 6.**). Among these participants, 220 agreed to engage with the modelling questionnaire, resulting in a response rate of 77%. The main reasons for the students' dropout were lack of interest, fatigue from workshop activities, and desire to take a break, as a break was scheduled immediately after the completion of the modelling questionnaire. We also acknowledge the challenges associated with survey research, such as social desirability and peer influence, which might affect the reliability of responses.

The workshops took place in real-life settings during school hours, replacing compulsory social science and geography classes. The entire-day workshops consisted of three modules involving the themes of general knowledge of the green transition, exploring conflicts of interest, and climate justice, with each session followed by a questionnaire to capture pupil perspectives. The comprehensive process, from designing workshop materials and questionnaires to conducting the workshops and analysing pupil choices, is outlined in Ref. [25].

The majority of participating students were enrolled in the general study program, and only one class was enrolled in the vocational study program. The schools were located in three different counties, with two schools situated in more urban settings than the other three.

## Supplementary References